\documentclass[article,useAMS,usenatbib]{mnras}
\usepackage{txfonts}
\usepackage{natbib}
\usepackage{graphicx}
\usepackage{dcolumn}

\begin{document}

\title[Small-scale clustering of nano-dust grains]{Small-scale clustering of nano-dust grains in supersonic turbulence}

\author[Mattsson et al.]{L. Mattsson$^{1}$\thanks{E-mail: lars.mattsson@su.se},  
J.~P.~U.~Fynbo$^{2,3}$,
B. Villarroel$^{1,4}$\\
$^1$Nordita, KTH Royal Institute of Technology and Stockholm University, Roslagstullsbacken 23, SE-106 91 Stockholm, Sweden\\
$^2$The Cosmic Dawn Center (DAWN)\\
$^3$Niels Bohr Institute, Lyngbyvej 2, DK-2100 Copenhagen, Denmark\\
$^4$Instituto de Astrof\'isica de Canarias (IAC), E-38200 La Laguna, Tenerife, Spain}

\pagerange{\pageref{firstpage}--\pageref{lastpage}} \pubyear{2019}

\maketitle

\label{firstpage}

\date{\today}

\begin{abstract} 
We investigate the clustering and dynamics of nano-sized particles (nano-dust) in high-resolution ($1024^3$) simulations of compressible isothermal hydrodynamic turbulence. It is well-established that large grains will decouple from a turbulent gas flow, while small grains will tend to trace the motion of the gas. We demonstrate that nano-sized grains may cluster in a turbulent flow (fractal small-scale clustering), which increases the local grain density by at least a factor of a few. In combination with the fact that nano-dust grains may be abundant in general, and the increased interaction rate due to turbulent motions, aggregation involving nano dust may have a rather high probability.  Small-scale clustering will also affect extinction properties.  As an example we present an extinction model based on silicates, graphite and metallic iron, assuming strong clustering of grain sizes in the nanometre range, could explain the extreme and rapidly varying ultraviolet extinction in the host of GRB\,140506A.
\end{abstract}


\section{Introduction}
Nano-sized dust grains, like graphitic particles, polycyclic aromatic hydrocarbons (PAHs) or even nano diamonds, are by number the most abundant type of dust in the interstellar medium (ISM). Their properties are different from those of larger grains and bulk material; they have large proportion of surface atoms and their sizes are actually smaller than (or at least similar to) basic scales such as the Landau radius, Debye length, De Brooglie wavelength etc. and may therefore show unique optical and kinematic properties \citep{Li12}. These grains make up only a tiny fraction of the total interstellar dust mass, but are believed to be a significant source of extinction, in particular if the grains are uniformly and isotropically distributed in the ISM. But if they are concentrated into small clumps with essentially dust free regions in between, the net extinction is reduced. Whether nano dust is clustered or not is therefore important.  Moreover, nano-sized sp$^2$-bonded carbon dust has for decades been thought to to be the main carrier of the 2175 \AA~extinction feature \citep[see, e.g.,][]{Pei92,Draine03,Jones13}, which indicates that small-scale clustering of such grains may be an important phenomenon to consider in order to explain variations of the ultraviolet (UV) extinction curve in general. 

Small particles, like nano-dust grains, may form clustered structures on scales smaller than the typical length scale of a turbulent flow. Such {\it small-scale clustering}\footnote{When we refer to ``small scale'' in the present paper, we mean scales much smaller that the characteristic length scale of the flow, which are not small compared to, e.g., the particles.}, a.k.a. preferential concentration, of particles in turbulent flows is a classical problem in fluid mechanics and statistical physics.  For incompressible flows, numerous studies have shown that centrifuging of particles away from vortex cores leads to accumulation of particles in convergence zones \citep[see, e.g.,][]{Maxey87,Squires91,Eaton94,Bec05,Yavuz18}. Thus, vorticity and inertia of the particles are decisive for the amount of clustering \citep[see][and references therein]{Toschi09}. The resultant fractal clustering has been studied and simulated quite extensively \citep[see, e.g.,][]{Sundaram97,Hogan99,Hogan01,Bec03,Bec07,Bec07b,Bec10,Bhatnagar18}, while it is not well-known to what extent it occurs in compressible supersonic turbulence. The latter is of course the regime which is interesting from an astrophysical perspective. With density scales typical for interstellar environments, recent results \citep{Hopkins16,Mattsson19a} point at small-scale clustering reaching its maximum for grains in the nanometre range (radii $a = 1\dots 100$~nm). 

In astrophysics, studies of clustering of particles in proto-planetary discs are common, but these studies often do not go significantly below the Kolmogorov scale and may thus not reach the clustering regime of interest in the ISM \citep[see, e.g.,][]{Pan11,Pan13,Pan14c}. \citet{Downes12,Hopkins16,Mattsson19a} have presented simulations of the turbulent dynamics of an interstellar MC including a dust phase from which a consistent picture emerge; small grains (radii $\lesssim 0.1\,\mu$m) tend to cluster and follow the gas, while larger grains ($\gtrsim 1\,\mu$m) tend to decouple from the gas flow and not cluster notably. 
For particles tracing the gas, the density variance due to turbulence will lead to an increased rate of grain growth by accretion of molecules in MCs (often referred to as ``dust condensation'') because of the non-linear dependence on molecular-gas density \citep{Mattsson19prep}. But if the clustering is strong, the dust will be quite far from uniformly distributed within observable gas structures. Thus, grain-growth processes which is based on accretion of molecules (e.g., condensation of ices or chemical reactions on grain surfaces) are highly dependent on the local environment.  Similarly, the collision rate of dust particles show significant variance due to spatial clustering, leading to an enhanced growth by coagulation \citep{Zsom08,Ormel09}.

To put the above into context, it is important to emphasise that grain growth is an important dust-formation channel, not the least as a necessary replenishment mechanism to counteract dust destruction in the ISM \citep{McKee89,Draine90}. Abundance patterns in interstellar gas are consistent with dust depletion due to condensation in MCs \citep[see, e.g.,][]{Jenkins09,DeCia16}, and the fact that late-type galaxies seem to have steeper dust-to-gas gradients than metallicity gradients along their discs lend further support to this picture \citep{Mattsson12a,Mattsson12b,Mattsson14b,Vilchez19}.  

Nano-sized grains are typically {\it not} in thermal and radiative equilibrium with their surroundings, which creates a wide grain temperature distribution \citep{Purcell76,Dwek86,Draine03}. Thus, they may affect the infrared flux-to-mass ratio in the same way as a range of sizes would for large grains in equilibrium \citep{Mattsson15}. Compared to a homogeneous distribution, small-scale clustering may cause nano grains to emit more long-wavelength radiation relative to the overall extinction they cause. This, in turn, leads to underestimation of nano-dust abundance from observations.

The present paper aims to explore small-scale clustering of nano-sized dust grains by direct numerical simulations of hydrodynamic turbulence with Lagrangian inertial particles and discuss a few of its consequences. 

        \begin{table*}
  \begin{center}
  \caption{\label{simulations}  Basic properties and time-averaged physical parameters of the simulations. All simulation have the mean gas density and isothermal sound speed set to unity, i.e., $\langle\rho\rangle=c_{\rm s}=1$.}
  \begin{tabular}{l|llllllll}
  \hline
  \hline
  \rule[-0.2cm]{0mm}{0.8cm}
  & $f$  &$\langle\log(\rho_{\rm min})\rangle$ & $\langle\log(\rho_{\rm max})\rangle$ & $\mathcal{M}_{\rm rms}$ & $\mathcal{M}_{\rm max}$ &Re$/L$ & Drag law & Forcing type\\
  \hline
  Case I & $4.0$ & $-4.25\pm 0.66$ & $1.52\pm 0.09$ & $3.22\pm 0.11$& $9.58\pm 0.68$ & $206\pm 7$ & $\mathcal{W}\ll 1$ & compressive\\
  Case IIa & $4.0$ & $-4.62\pm 1.03$ & $1.53\pm 0.08$ & $3.28\pm 0.11$& $10.03\pm 1.05$ & $210\pm 7$ & All $\mathcal{W}$ & compressive\\
  Case IIb & $4.0$ & $-3.17\pm 0.43$ & $1.32\pm 0.05$ & $3.56\pm 0.07$& $9.83\pm 0.51$ & $228\pm 5$ & All $\mathcal{W}$ & solenoidal\\ 
    Case IIIa & $4.0$ & $-4.24\pm 0.60$ & $1.51\pm 0.07$ & $3.17\pm 0.15$& $9.50\pm  0.86$ & $203\pm 9$ & -- & compressive\\
  Case IIIb & $4.0$ & $-3.32\pm 0.53$ & $1.32\pm 0.06$ & $3.57\pm 0.10$& $9.70\pm 0.54$ & $228\pm 7$ & -- & solenoidal\\ 
  \hline
  \hline
  \end{tabular}
  \end{center}
  \end{table*}

\begin{figure*}
      \resizebox{\hsize}{!}{
      \includegraphics{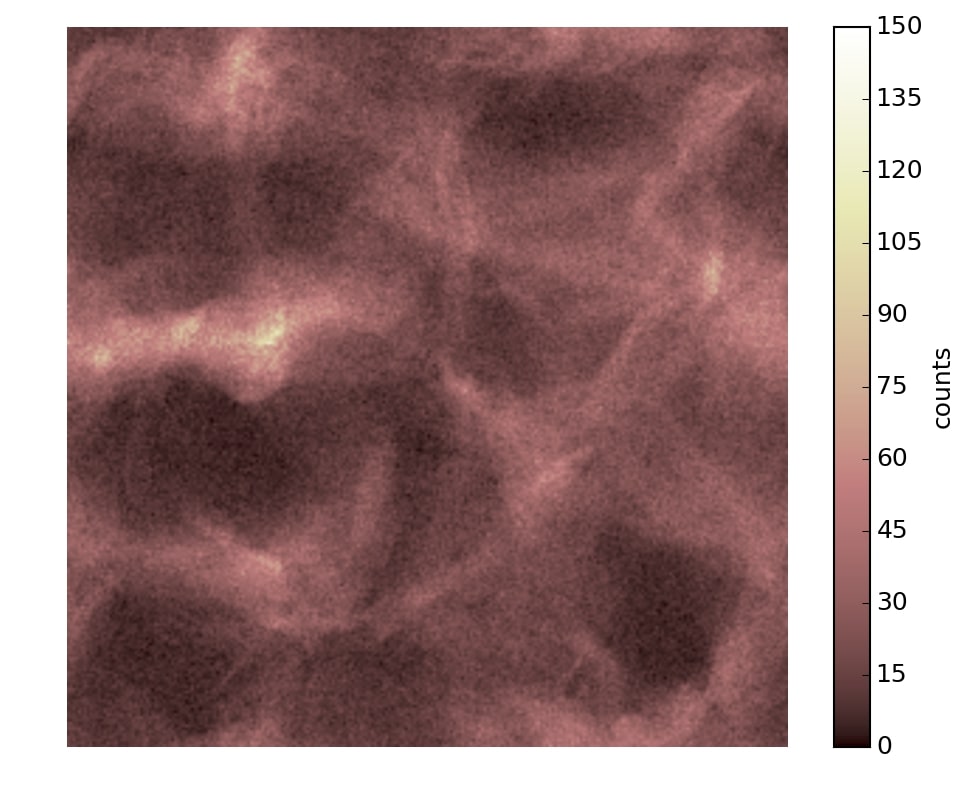}
      \includegraphics{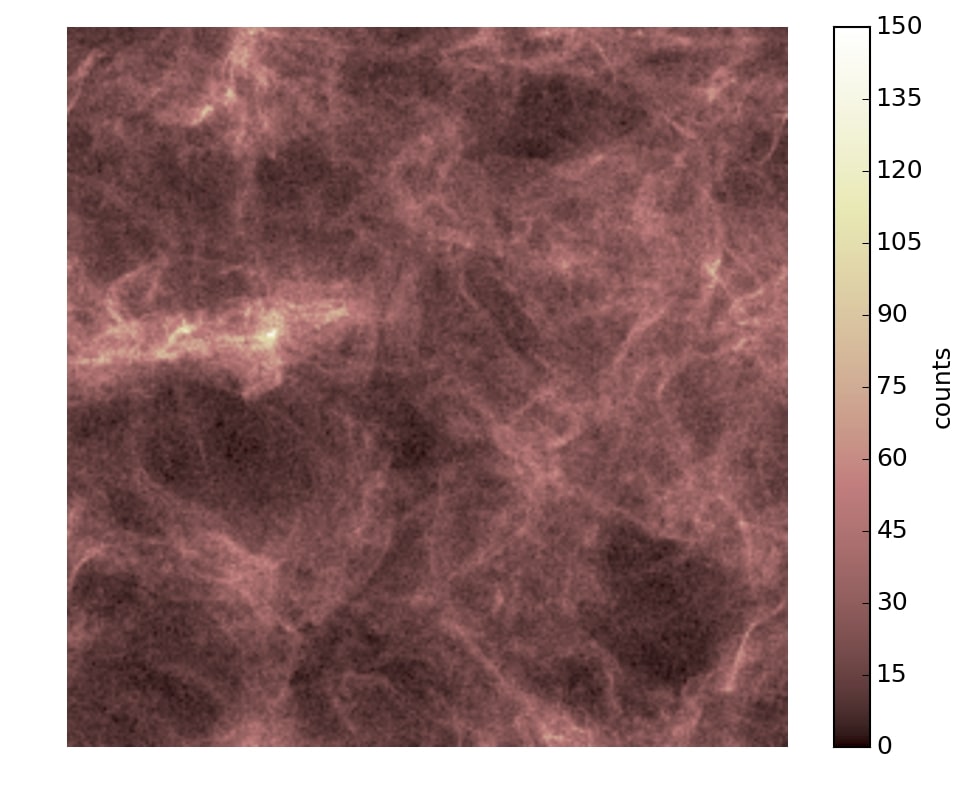}
      \includegraphics{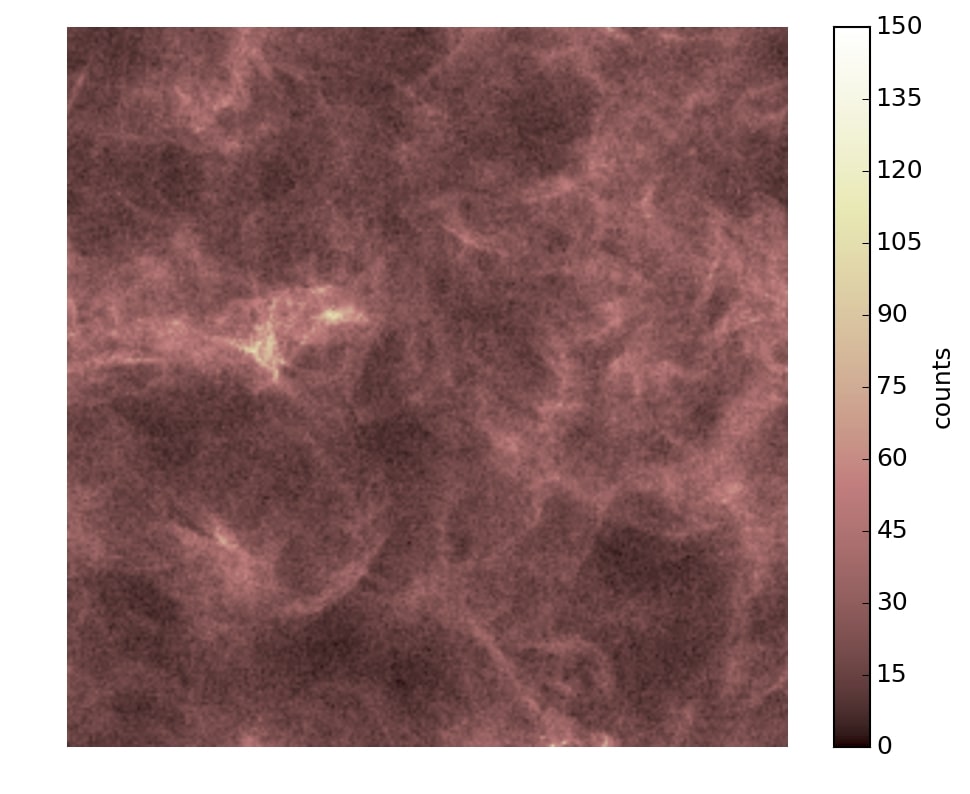}
      }
      \resizebox{\hsize}{!}{
      \includegraphics{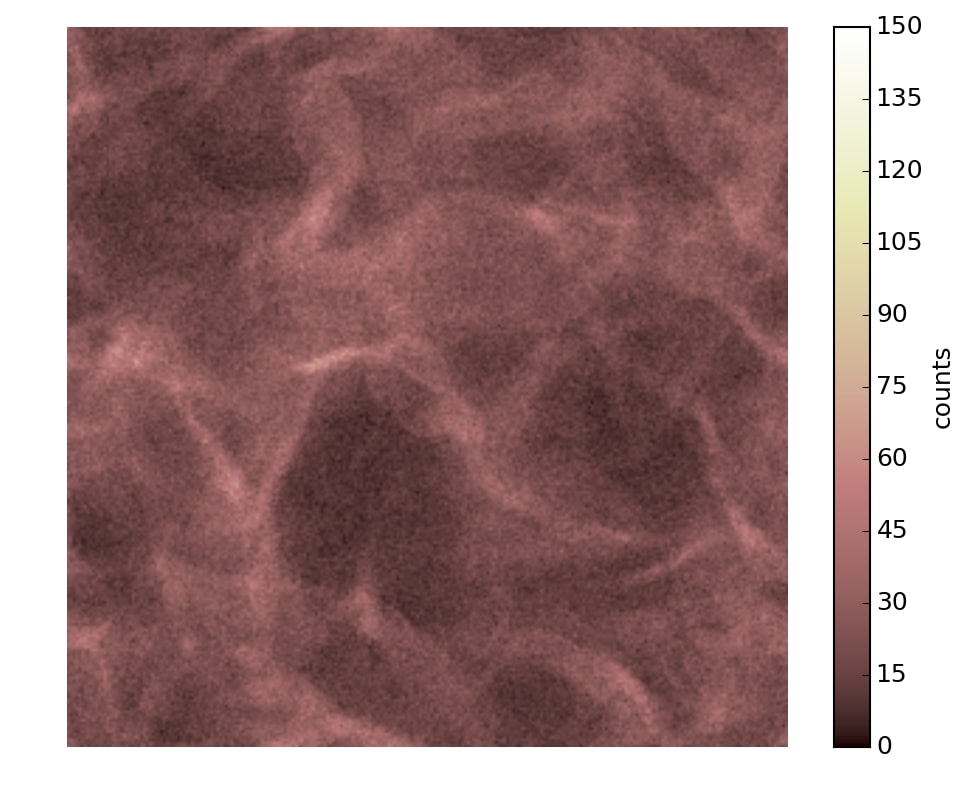}
      \includegraphics{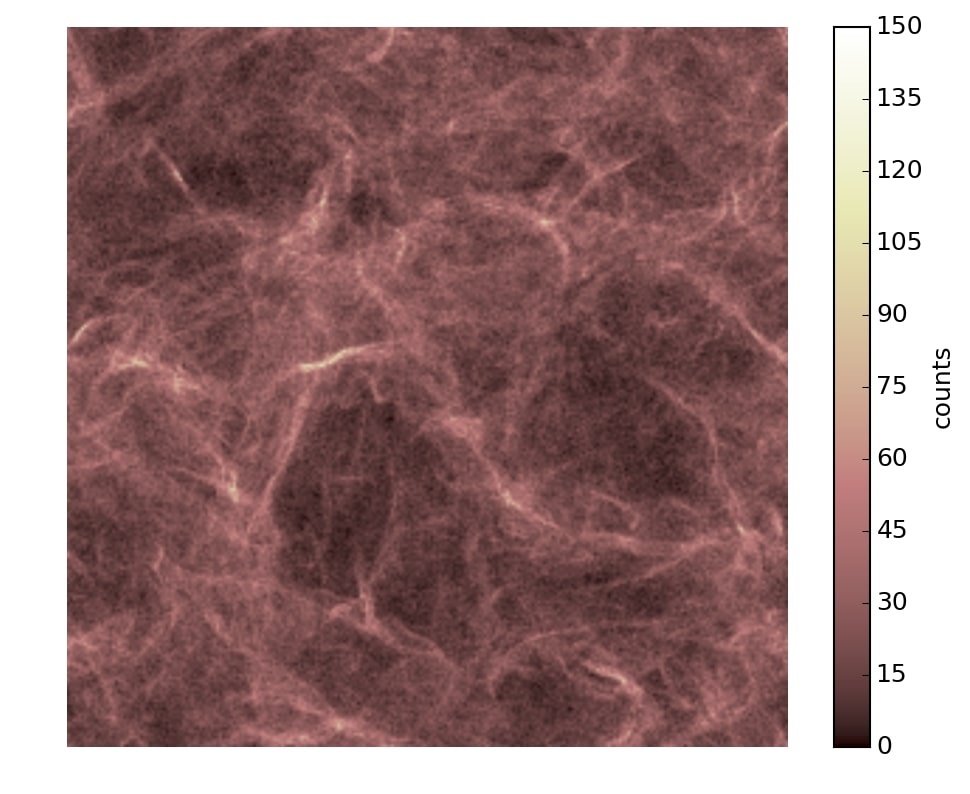}
      \includegraphics{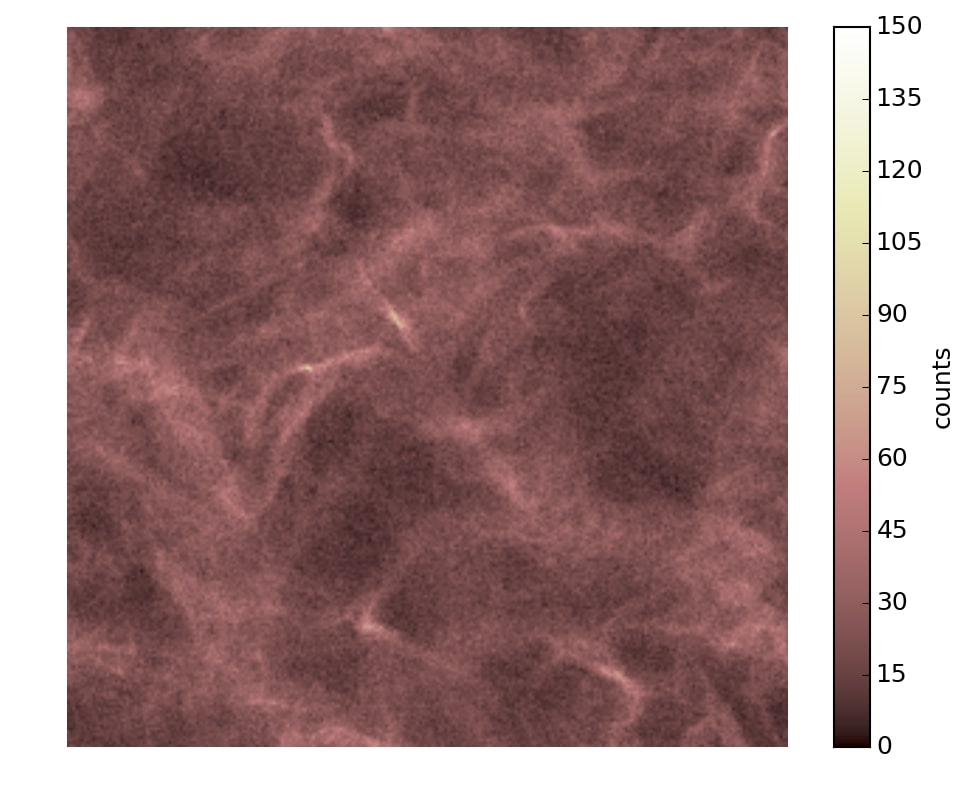}
      }
  \caption{\label{projections} Projected number density of dust grains. From left to right: smallest (inertial) nano grains in the simulation ($\alpha = 0.0017$, $a \approx 1$~nm), maximally clustered grains ($\alpha = 0.033$, $a \approx 25$~nm) and largest nano grains ($\alpha = 0.13$, $a \approx 100$~nm). Upper panels show the run with compressive forcing, while the lower panels show the run with solenoidal forcing.}
  \end{figure*}

\begin{figure*}
      \resizebox{\hsize}{!}{
      \includegraphics[width=61.8cm]{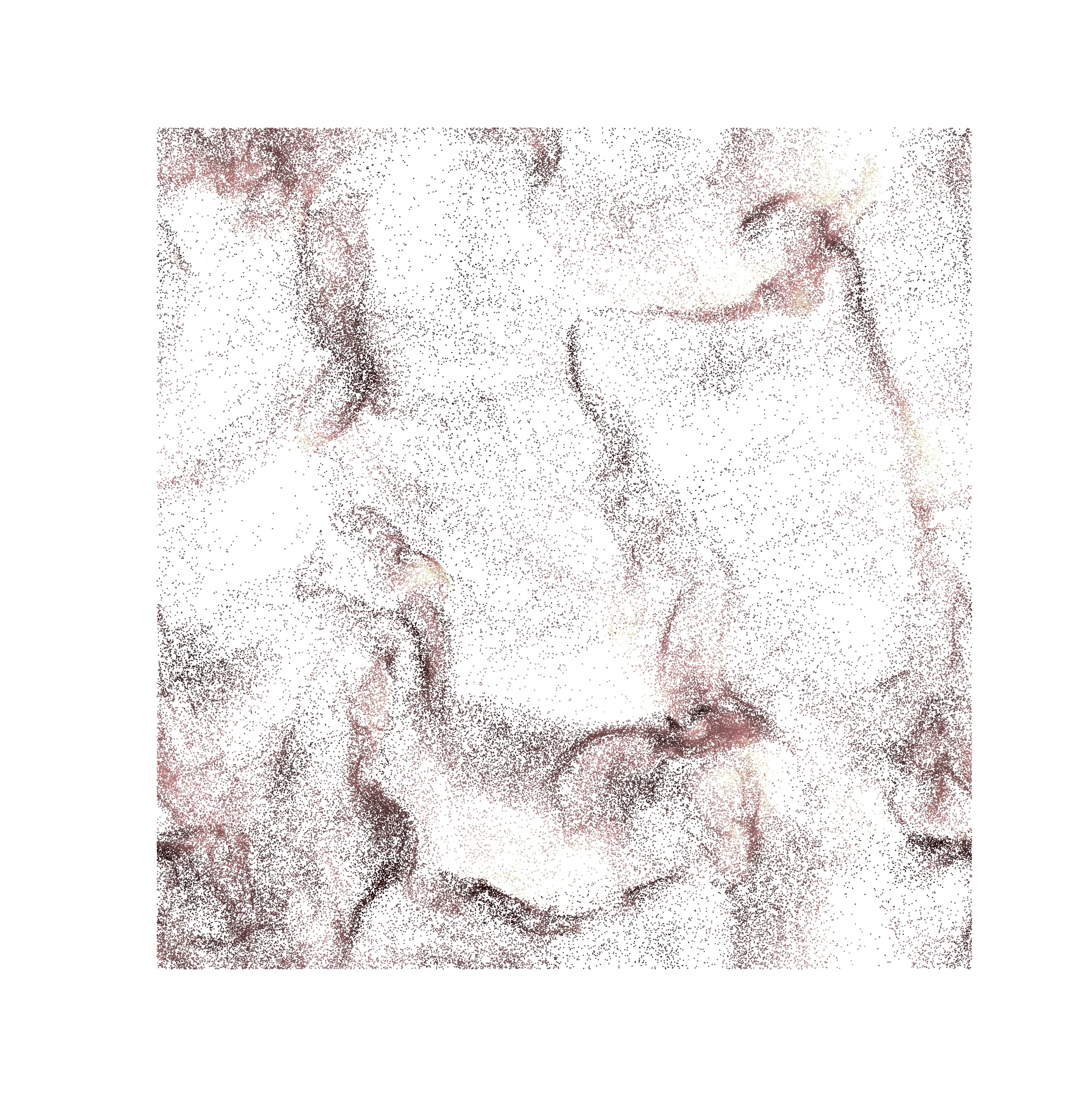}
      \includegraphics{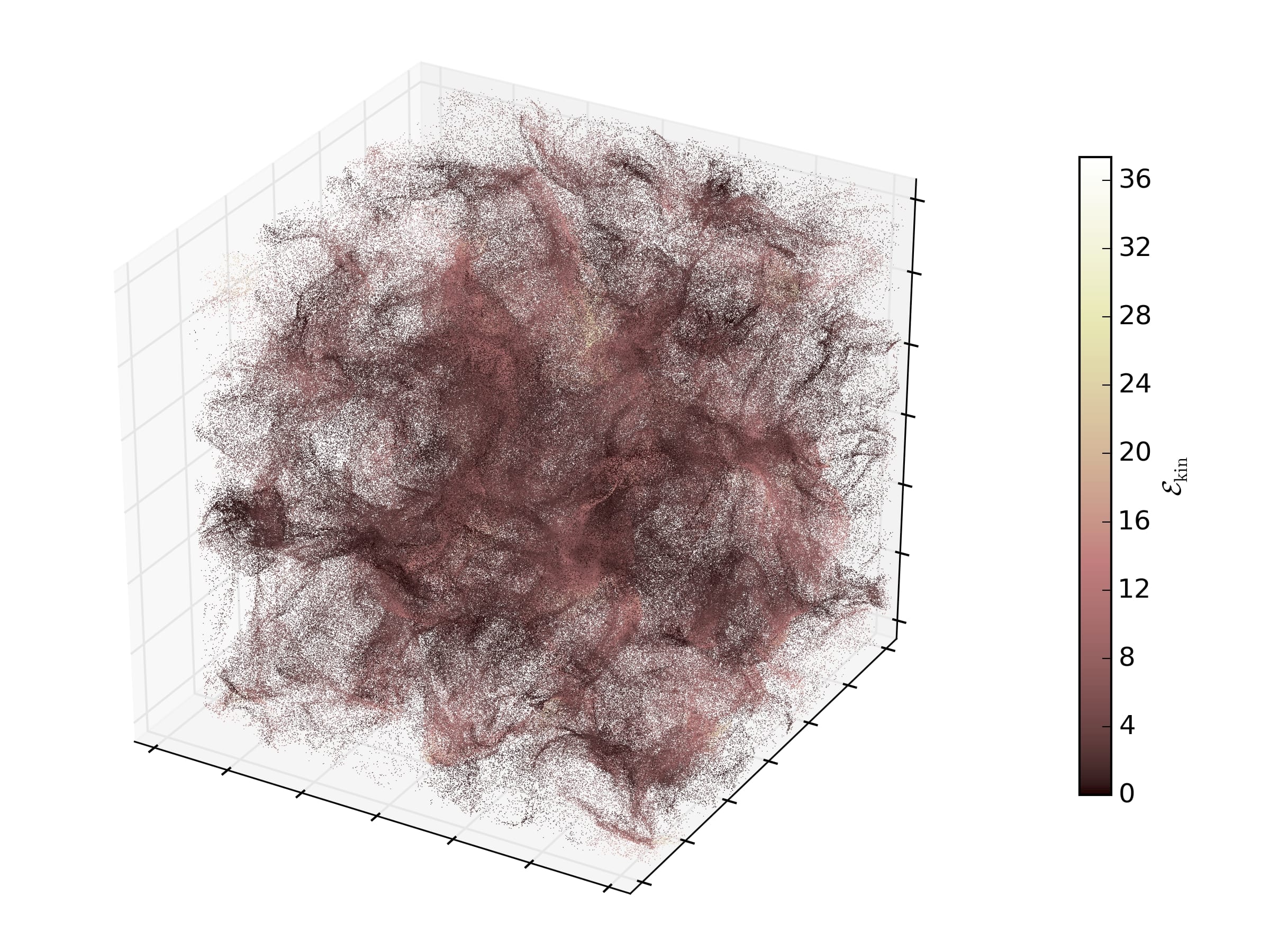}      
      }
  \caption{\label{3Ddust_cmp} Left: distribution of dust particles in a slice of thickness $\Delta L/L = 1/64$ taken through the middle of the simulation box. Right: 3D snapshot from the simulation with \textit{compressive forcing}, showing the dust density distribution of maximally clustered grains  ($\alpha = 0.033$, $a \approx 25$~nm). Colour coding indicate specific kinetic energy.}
  \end{figure*}        
      
\begin{figure*}
      \resizebox{\hsize}{!}{
      \includegraphics[width=61.8cm]{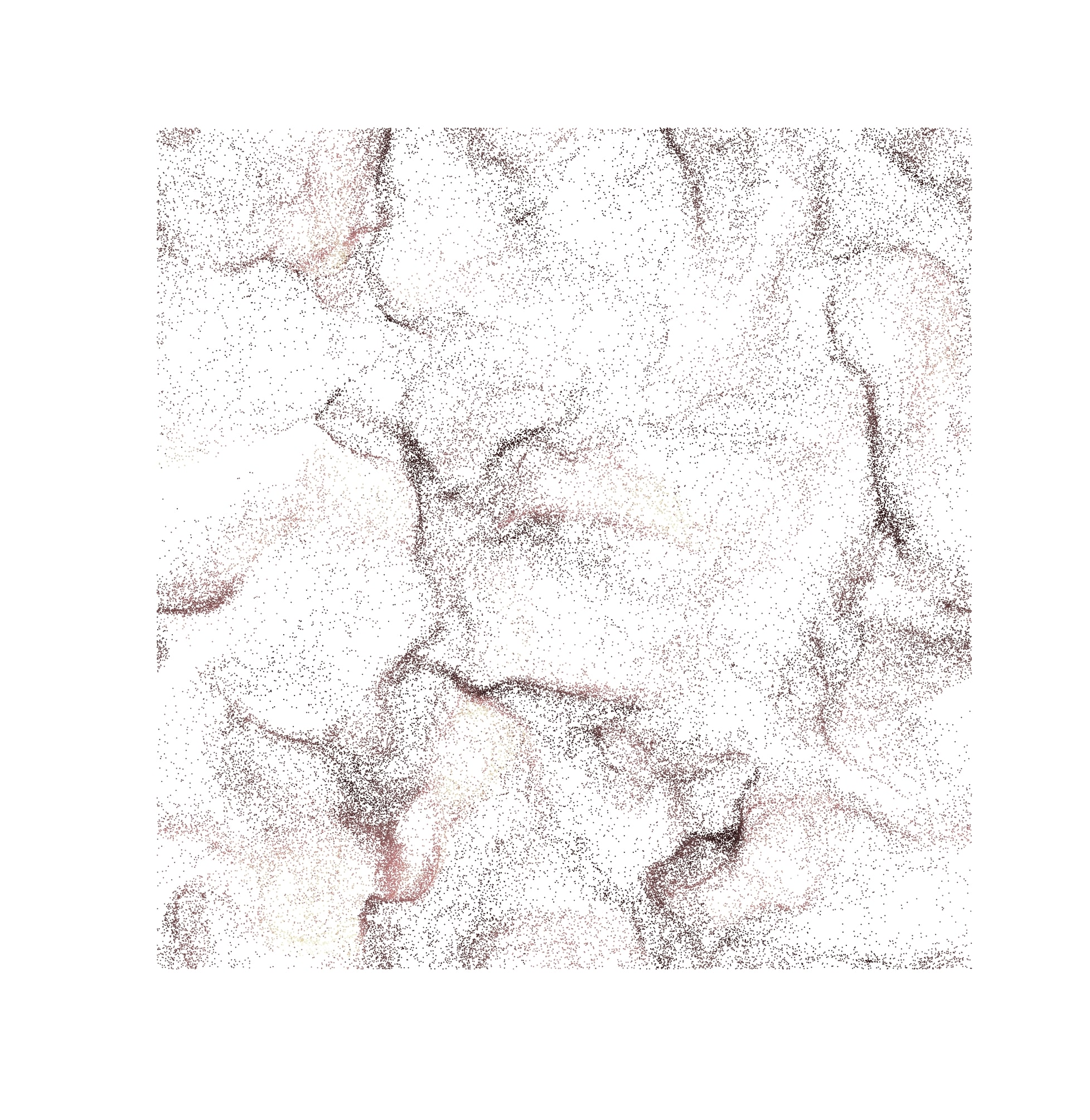}
      \includegraphics{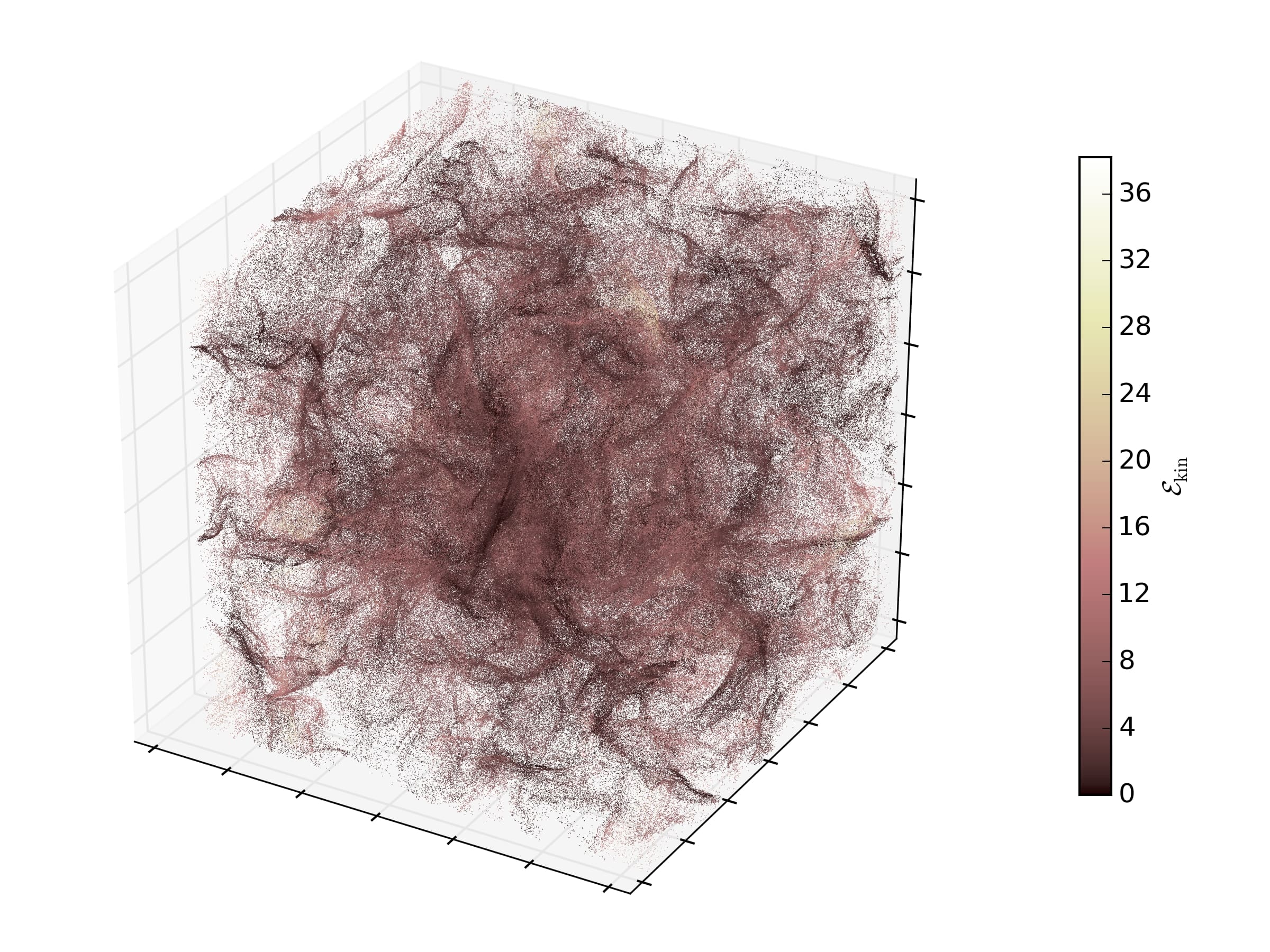}
      }
  \caption{\label{3Ddust_sol} Left: distribution of dust particles in a slice of thickness $\Delta L/L = 1/64$ taken through the middle of the simulation box. Right: 3D snapshot from the simulation with \textit{solenoidal forcing}, showing the dust density distribution of maximally clustered grains  ($\alpha = 0.033$, $a \approx 25$~nm). Colour coding indicate specific kinetic energy.}
  \end{figure*}

\section{Theory and methods}
\subsection{Simulation setup}
We model turbulent gas aimang at describing the interior of a molecular cloud (MC). We set up a high-resolution ($1024^3$) three-dimensional (3D) periodic-boundary box with sides of equal length $L = 2\upi$ and solve the standard hydrodynamic equations as described in, e.g., \citet{Mattsson19a}: the momentum equation, the continuity equation and an isothermal condition added as a closure relation, i.e., a constant sound speed $c_{\rm s} = 1$ (=~unit velocity). Dust particles are included as inertial particles in 15 size bins with $10^6 - 10^7$ particles in each. We use the {\sc Pencil Code}, which is a non-conservative, high-order, finite-difference code (sixth order in space and third order in time). For a more detailed description of the code, see \citet{Brandenburg02}. For a more detailed description of the type of turbulence simulations we employ in this study we refer to \citet{Mattsson19a}.

\subsection{External forcing}
To maintain supersonic steady-state turbulence we need some kind of external forcing.
We therefore include a white-in-time stochastic forcing term with both solenoidal (rotational) and compressible components. The forcing is applied at low wave-numbers in Fourier space, where the forcing is a  stochastic process integrated using the Euler-Maruyama method. In the present study we consider two types of forcing; either purely compressive or purely solenoidal forcing (see Table \ref{simulations}) in order to explore if there will be any qualitative differences in the dust dynamics. It is known that the resultant gas-density PDFs and the fractal properties of the gas are significantly different depending on whether the forcing is compressive or solenoidal \citep{Federrath09,Federrath10}. 
 
\subsubsection{Kinetic drag force}
\label{dustflow}
Inertial particles suspended in a gaseous medium will show delayed response to kinetic drag from gas particles.  Assuming the interstellar dust is accelerated by a turbulent gas flow via an \citet{Epstein24} drag law, the equation of motion for dust particles is
\begin{equation}
\label{stokeseq}
{{\rm d} \mathbfit{v}\over {\rm d} t}  = {\mathbfit{u}-\mathbfit{v}\over \tau_{\rm s}},
\end{equation}
where $\mathbfit{v}$ and $\mathbfit{u}$ is the velocities of dust and gas, respectively, and $\tau_{\rm s}$ is the stopping time, i.e., the timescale of acceleration (or deceleration) of the grains. $\tau_{\rm s}$ in the Epstein limit depends on the grain radius $a$ and bulk material density $\rho_{\rm gr}$ well as the gas density $\rho$ and the relative Mach number $\mathcal{W}_{\rm s} = |\mathbfit{u}-\mathbfit{v}|/c_{\rm s}$ \citep{Schaaf63}. In the limit $\mathcal{W}_{\rm s}\ll 1$, we obtain 
\begin{equation}
\label{stoppingtime_incomp}
\tau_{\rm s}(\mathcal{W}_{\rm s}\ll 1) =  \sqrt{\upi\over 8}{\rho_{\rm gr}\over\rho}{a\over  c_{\rm s}} \equiv \tau_{\rm s,\,0},
\end{equation}
where $c_{\rm s}$ has replaced the thermal mean speed of molecules. The $\mathcal{W}_{\rm s}\ll 1$ case typically corresponds to small sonic Mach numbers, i.e., $\mathcal{M}_{\rm s}\ll 1$. For large $\mathcal{M}_{\rm s}$, we expect $\mathcal{W}_{\rm s}\gg 1$ and
\begin{equation}
\label{stoppingtime_comp}
\tau_{\rm s} (\mathcal{W}_{\rm s}\gg 1)=  {4\over 3}{\rho_{\rm gr}\over\rho}{a\over  |\mathbfit{u}-\mathbfit{v}|}.
\end{equation}
Combining these two limits, we then obtain a convenient formula which is sufficiently accurate for our purposes \citep{Kwok75,Draine79},
\begin{equation}
\label{stoppingtime}
\tau_{\rm s} = \tau_{\rm s,\,0} \left(1 + {9\upi\over 128}{|\mathbfit{u}-\mathbfit{v}|^2\over c_{\rm s}^2 } \right)^{-1/2}.
\end{equation}
The second term inside the parentesis can be seen as a correction for supersonic flow velocities and compression.

Non-inertial particles, a.k.a. tracer particles, will have $\mathbfit{v} = \mathbfit{u}$ and be position coupled with the medium in which they reside. It is often assumed that the approximation $\mathbfit{v} \approx \mathbfit{u}$ is justified when the stopping time is much shorter than the characteristic timescale of the flow, which is the case for very small interstellar dust particles. The smallest nano dust (typically $a\sim 1$~nm) would be an example of such dust particles. However, as the ISM is highly compressible, with high $\mathcal{M}_{\rm s}$, even a very small amount of inertia can be important. A similar approximation, but perhaps better than $\mathbfit{v} \approx \mathbfit{u}$, could be assuming that $\mathcal{W}_{\rm s} \ll 1$. Any difference between clustering of tiny particles in incompressible (or nearly incompressible) turbulence and supersonic compressible turbulence is then due to compressibility and the variance of $\rho$. If this hypothesis is correct, the clustering of nano dust should be essentially unaffected by the correction for large $\mathcal{W}_{\rm s}$ described above.

We will use two prescriptions for kinetic Epstein-type drag and in total consider five cases:
\begin{itemize}
\item {\bf Case I:}~$\tau_{\rm s}$ prescription for low $\mathcal{W}$ (Eq. \ref{stoppingtime_incomp}) and turbulence induced by purely compressive forcing.
\item {\bf Case IIa:}~$\tau_{\rm s}$ prescription including correction for large $\mathcal{W}$ (Eq. \ref{stoppingtime}) and purely compressive forcing.
\item {\bf Case IIb:}~Same as Case IIa, but solenoidal instead of compressive forcing. 
\item {\bf Case IIIa:}~non-inertial (``tracer'') particles and turbulence induced by purely compressive forcing.
\item {\bf Case IIIb:}~Same as Case IIIa, but solenoidal instead of compressive forcing. 
\end{itemize}
The two most realistic cases are IIa and IIb and these two will get some extra attention in the analysis below.

\subsubsection{Notes on other forces on dust grains}
Nano-sized grains are easily affected by other types of external forces such as magnetic fields and radiation pressure. If electrically charged, magnetic fields and ionised gas (protons) will exert forces on the grains, which do not necessarily act in the same direction as the kinetic drag from neutral gas. If exposed to a radiation field, the grains may also experience a systematic directional acceleration due to momentum transfer from photons to dust grains (``radiation pressure''). Thus, the dynamics and clustering of nano-dust grains can be very complex.

However, as opposed to nano dust in, e.g., the solar wind, where radiation, plasma drag and magnetic fields play crucial roles \citep[see, e.g.,][]{Czechowski18}, cold MCs in the ISM represent an environment which allows for a significantly simplified description. Before the onset of star formation in an MC, there is little radiation that can reach the inner parts, where most of the dust is located. Without a significant radiation field, the gas will maintain a very low degree of ionisation, radiation pressure will be minimal, grains will not carry much electrical charge and thus be mostly unaffected by magnetic fields or Lorentz forces. In the present study, we simulate the conditions in a starless  MC, where it is reasonable to assume that kinetic gas drag is the dominant force acting upon the grains. As soon as star formation sets in the conditions will be very different, but we leave that for a future study.

\subsubsection{The grain-size parameter}
We employ a ``grain-size parameter'',
\begin{equation}
\alpha = {\rho_{\rm gr}\over\langle \rho\rangle}{a\over L} ,
\end{equation}
which is the parameter used  by \citet{Hopkins16,Mattsson19a}. Because the total gas mass $M = \langle\rho\rangle\,L^3$, as well as $\rho_{\rm gr}$ and $a$ are constants, $\alpha$ must also be a constant. It is known from studies of incompressible turbulence that  particles of different Stokes number ${\rm St}$ can be maximally concentrated on different length scales \citep{Bec03, Bec07b, Zaichik03}. To some extent this may be a reflection of the fact that ${\rm St}$, as it is usually defined for incompressible flows, depends on the dissipative scale of the carrier flow \citep{Hogan99}, which is not independent of resolution\footnote{In our simulations, the particles are not treated on a mesh, but the mesh used for the carrier flow (the gas) still imposes an indirect resolution limit.} and forcing scale, and particularly so when the flow is highly compressible and ${\rm St}$ is not even a universal number on average \citep{Mattsson19a}.  The disadvantage of ${\rm St}$ is that a highly compressible gas with ${\rm Re} \gg 1$ flowing around an object is never a Stokes flow. The Stokes number ${\rm St}$ can indeed be used even if  ${\rm Re} \gg 1$, but if $\mathcal{M}_{\rm rms}\gg 1$ too, ${\rm St}$ is not adequately well-defined. Therefore $\alpha$ may be a better dimensionless measure of grain size than  $\langle {\rm St}\rangle$ for supersonic compressible flows. However, $\alpha$ must still be anchored to the physical characteristics of the flow. Following \citet{Hopkins16} we can relate $L$ to the sonic length,
\begin{equation}
R_{\rm s} = R_{\rm MC}\,{c_{\rm s}^2\over |\mathbfit{u}\cdot\mathbfit{u}|} = {R_{\rm MC}\over \mathcal{M}_{\rm rms}^2},
\end{equation}
where $R_{\rm MC}$ is the effective radius of the modelled MC (or part thereof). In the present case we can assume $R_{\rm MC}\sim L$. 

The largest clumps within an MC have characteristic radii $\sim 1$~pc \citep[see, e.g.,][]{Loren89}. With $\mathcal{M}_{\rm rms} \sim 3$, as in the present work, we find that $R_{\rm s} \sim 0.1$~pc, which is consistent with the $R_{\rm s}$ estimated in Milky Way-like GMCs and the empiric linewidth-size relation $\mathcal{M}_{\rm rms} \sim (R/R_{\rm s})^{1/2}$.  The resultant scaling relation is
\begin{equation}
\label{phys_L}
L \sim  10\,{\rm pc}\,\left({\mathcal{M}_{\rm rms}\over 10} \right)^{2} \left({R_{\rm s}\over 0.1\,{\rm pc}} \right).
\end{equation}
Again following \citet{Hopkins16} the physical size of the grains can be obtained from
\begin{equation}
\label{phys_a}
a = 0.4\,\alpha\, \left({L\over 10\,{\rm pc}} \right)\left({\langle n_{\rm gas}\rangle \over 10\,{\rm cm}^{-3}} \right) \left({\rho_{\rm gr}\over 2.4\,{\rm g\,cm}^{-3}} \right)^{-1}\,\mu{\rm m},
\end{equation}
where $\langle n_{\rm gas}\rangle$ is the average number density of gas particles (molecules). The physical scales introduced in eq. (\ref{phys_a}) represent the typical size of large MCs, the characteristic $\langle n_{\rm gas}\rangle$ in the cold-phase ISM and the average $\rho_{\rm gr}$ of Galactic interstellar dust. Combining eqns. (\ref{phys_L}) and (\ref{phys_a}) gives\footnote{An alternative definition of the ``grain-size parameter'' (which takes the characteristics of the flow into account) could be
\begin{equation}
\tilde{\alpha} = {1\over\mathcal{M}_{\rm rms}^2} {\rho_{\rm gr}\over\langle \rho\rangle}{a\over R_{\rm s}},
\end{equation}
but we stick to the definition by \citet{Hopkins16} to avoid confusion when comparing results.}, 
\begin{equation}
\label{phys_a_2}
a \sim \alpha\, \left({\mathcal{M}_{\rm rms}\over 10} \right)^{2} \left({R_{\rm s}\over 0.1\,{\rm pc}} \right) \left({\langle n_{\rm gas}\rangle \over 10\,{\rm cm}^{-3}} \right) \left({\rho_{\rm gr}\over 2.4\,{\rm g\,cm}^{-3}} \right)^{-1}\,\mu{\rm m}.
\end{equation}
Assuming values of $R_{\rm s}$, $\langle n_{\rm gas}\rangle$ and $\rho_{\rm gr}$ typical for MCs, we have $a\sim \alpha\,\mu$m. We use a factor of $0.75\,\mu$m when converting from $\alpha$ to $a$ (our standard scaling) but this number of course depends on the assumed properties. The smallest particles in our simulations have $\alpha = 0.0016$ and the largest $\alpha = 1.6$, which corresponds to $a \approx 1.2$~nm and $a \approx 1.2\,\mu$m, respectively. (Nano-dust is usually defined as grains with radii $a = 1\dots 100$~nm.)

\subsubsection{Small-scale clustering}
One of the main objectives of the present study is to find the maximum degree of clustering and at which $\alpha$ this maximum occurs. Centrifuging of particles away from vortex cores leads to accumulation of particles in convergence zones which is different from the increased number density of dust grains $n_{\rm d}$ due to compression of the gas and dust \citep{Maxey87, Squires91, Eaton94, Bec05, Pumir16, Yavuz18}. 

We can quantify clustering of grains using nearest neighbour statistics (NNS), obtained with the $kd$-tree algorithm \citep{Bentley75} including edge corrections. We can then indirectly determine the correlation dimension $d_2$, a kind of fractal dimension defined as $d_2 = d\ln \mathcal{N}/d\ln r$ as $r\to 0$, where $\mathcal{N}$ is the number of particles (grains) surrounding a reference particle within a ball of radius $r$ \citep{Monchaux12}. Thus, $d_2$ can be determined from $g(r)$, the radial distribution function (RDF), which is related to the two-point correlation function $\xi(r) = g(r) - 1$  \citep{Martinez03}. In 3D, the cumulative distribution $H(r)$ of the first nearest neighbour distances (1-NNDs) is in turn related to $g(r)$. More precisely,
\begin{equation}
H(r) = \tilde{n}g(r) \exp\left[-\int_0^r \tilde{n}\,g(s)\,4\pi\,s^2\,ds \right],
\end{equation}
where $\tilde{n}$ is the renormalised average number density \citep{Torquato90,Bhattacharjee03}.
For a homogeneous Poisson process $g=1$, so that $H(r)$ reduces to the exact result discussed by, e.g., \citet{Chandrasekhar43}. Clearly, $H(r) \to \tilde{n} g(r)$ as $r\to 0$. In this limit $g(r) \propto r^{d_2 -3}$ and therefore the $h(r) = dH(r)/dr \propto r^{d_2-1}$. We can thus determine $d_2$ from the NNS by computing $h(r)$ or $H(r)$, where the latter is obtained by rank-order techniques, which minimises the error.  The agreement between this method and the more direct metod used by \citet{Bhatnagar18} is excellent.

A major advantage with the NNS method is that it does not require explicit use of a binning radius $\delta r_{\rm bin}$, which in such a case should be significantly larger than the distance $\delta r_{\rm min}$ between the two closest particles. With $h(r)$ or $H(r)$, we can determine $g(r,\alpha)$ or $\xi(r,\alpha)$ for small $r$ and a given $\alpha$. To explore $g(r,\alpha)$ as a function $\alpha$ only, we must chose an evaluation radius $r_{\rm e}$, which is sometimes defined in terms of the Kolmogorov scale $\eta$ \citep[see, e.g.,][]{Pan11}. This is reasonable for $\mathcal{M}_{\rm rms} \lesssim 1$, but since we consider $\mathcal{M}_{\rm rms} > 3$, we will choose $r_{\rm e}$ to some fraction of the average 1-NND $\langle r \rangle$. To ensure that we evaluate $g(r,\alpha)$ in the power-law regime for small $r$, we must have $r_{\rm e} \lesssim {1\over 2} \langle r\rangle$ and to stay above the resolution limit of the mesh, $r_{\rm e} \gtrsim {1\over 16} \langle r\rangle$. (Note that ${1\over 16} \langle r\rangle < \delta r_{\rm min}$ in most cases.) If there is no small-scale clustering, then $g(r_{\rm e}, \alpha) = 1$; if there is, then $g(r_{\rm e}, \alpha) > 1$.

\subsection{Simple radiative transfer}
\label{sec:rte}
Dust extinction depends on how the dust is distributed and thus also the degree of clustering. We consider a simplified radiative transfer (RT) problem where we omit absorption and emission from the gas and just calculate the effects on incident light due to dust. The RT equation (RTE) along a column/ray of light then simplifies to,
\begin{equation}
\label{rte}
{dI_\lambda\over ds} = -\rho(s)\,\kappa_{{\rm d},\,\lambda}(a,s)\,I_\lambda (s),
\end{equation}
where we can also assume that the dust opacity $\kappa_{{\rm d},\,\lambda}$ is due to pure absorption, since scattering is typically negligible for $a\ll \lambda/2\upi$. Then,
\begin{equation}
\label{dopac}
\rho(s)\,\kappa_{{\rm d},\,\lambda}(a,s) = \upi \int_0^\infty a^2\,n_{\rm d}(a,s)\,Q_{\rm ext}(a,\lambda)\,da,
\end{equation}
and the formal solution to (\ref{rte}) can be written
\begin{equation}
I_{\lambda}(s) = I_{\lambda}(0)\exp\left[-\upi \int_0^\infty a^2\,n_{\rm d}(a,s)\,Q_{\rm ext}(a,\lambda)\,da\right],
\end{equation}
where $Q_{\rm ext}$ is the ratio between the effective extinction cross-section and the geometric cross-section ($\upi\,a^2$) and $I_{\lambda}(0)$ is the intensity of the incident light. 
The RT for a plane-parallell case (distant radiation source) can then be calculated by numerical integration over one geometric dimension of the simulation box.

  \begin{figure}
      \resizebox{\hsize}{!}{
      \includegraphics{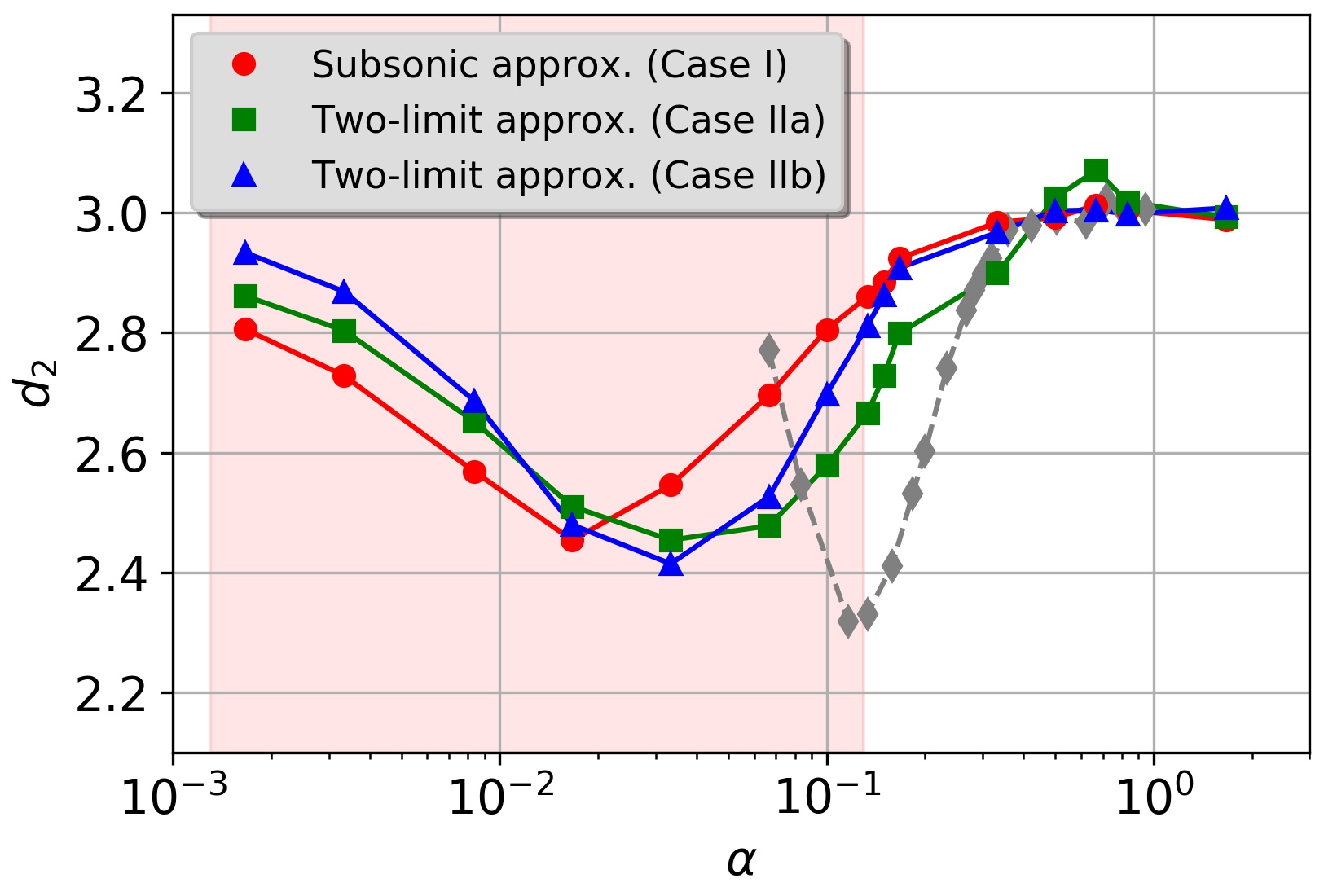}}
  \caption{\label{D2} Correlation dimension $d_2$ as function of the grain-size parameter $\alpha$. The grey diamonds connected by a dashed line in the background show the results from a simulation of incompressible turbulence by \citet{Bhatnagar18}. The simulation presented here (see legend) all show a much wider dip and minimum in $d_2$ at a lower $\alpha$. The shaded area marks the nano dust range for a typical physical scaling of the simulations.}
  \end{figure}  
  
  \begin{figure}
      \resizebox{\hsize}{!}{
      \includegraphics{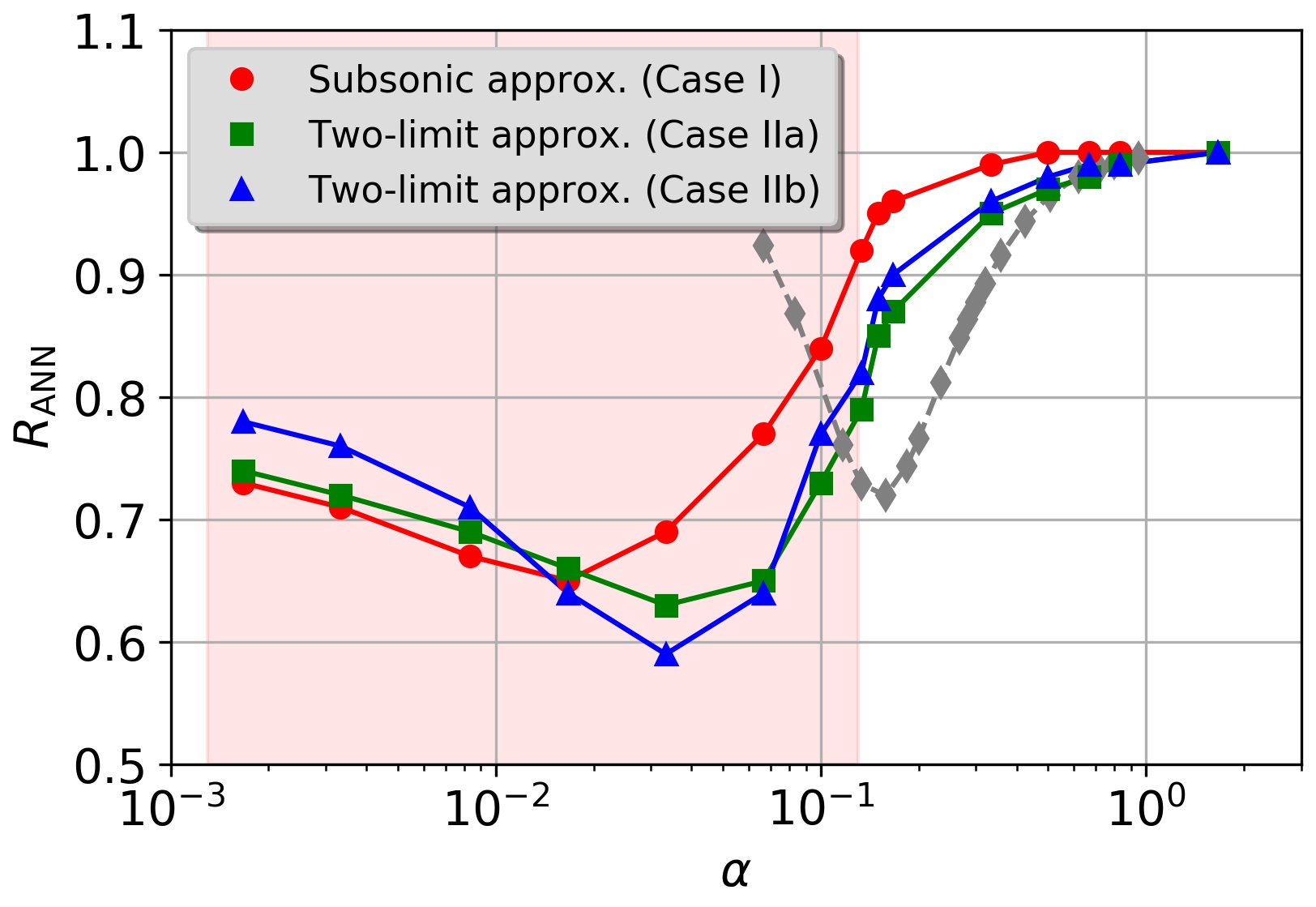}}
  \caption{\label{RANN} Average nearest neighbour distance ratio $R_{\rm ANN}$ as function of the grain-size parameter $\alpha$. As opposed to the case of incompressible turbulence, very small particles in compressible supersonic turbulence does not show $R_{\rm ANN}\to1$ as $\alpha\to0$, which is merely an effect of the compressibility of the medium. }
  \end{figure}  
  
    \begin{figure*}
      \resizebox{\hsize}{!}{
      \includegraphics{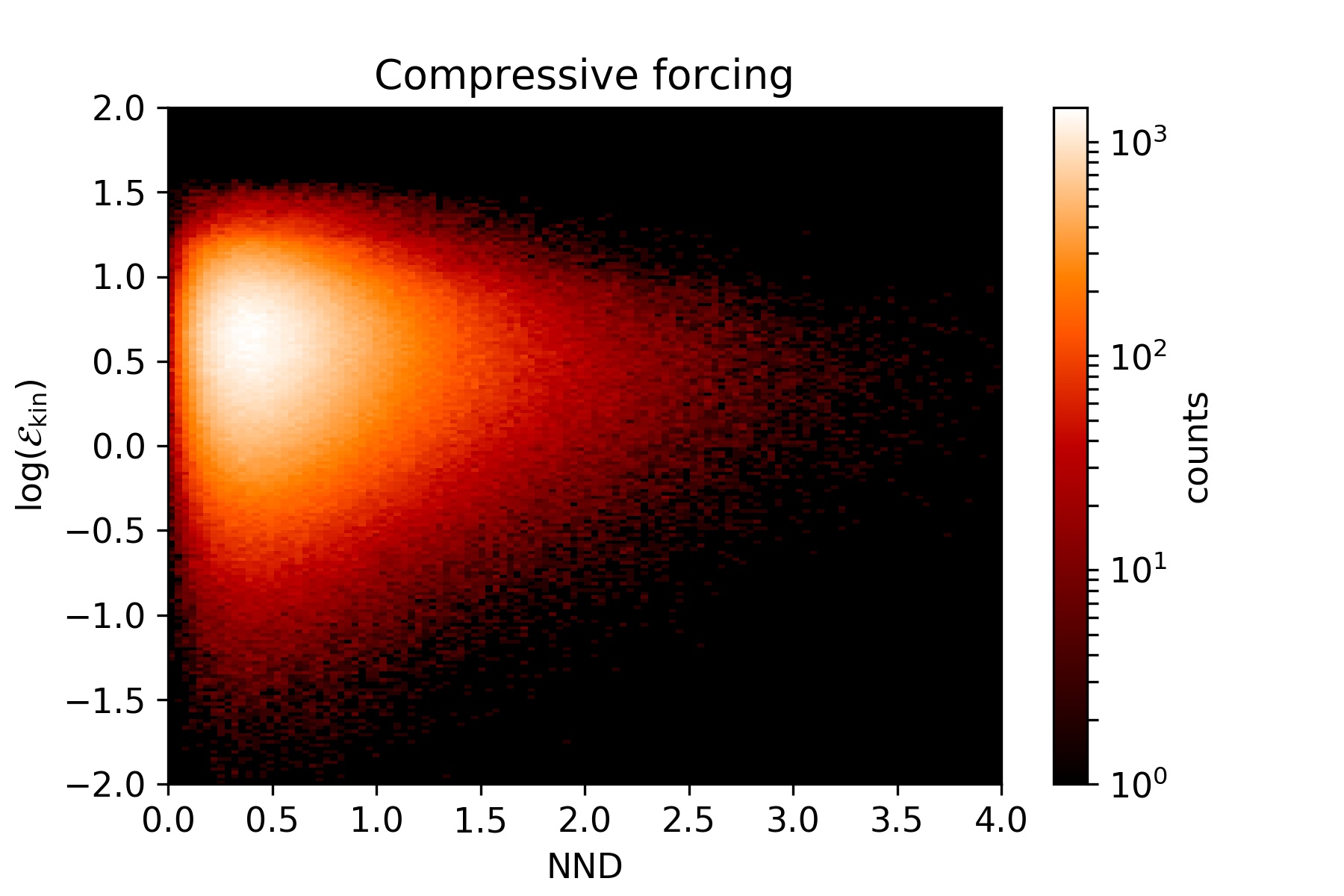}
      \includegraphics{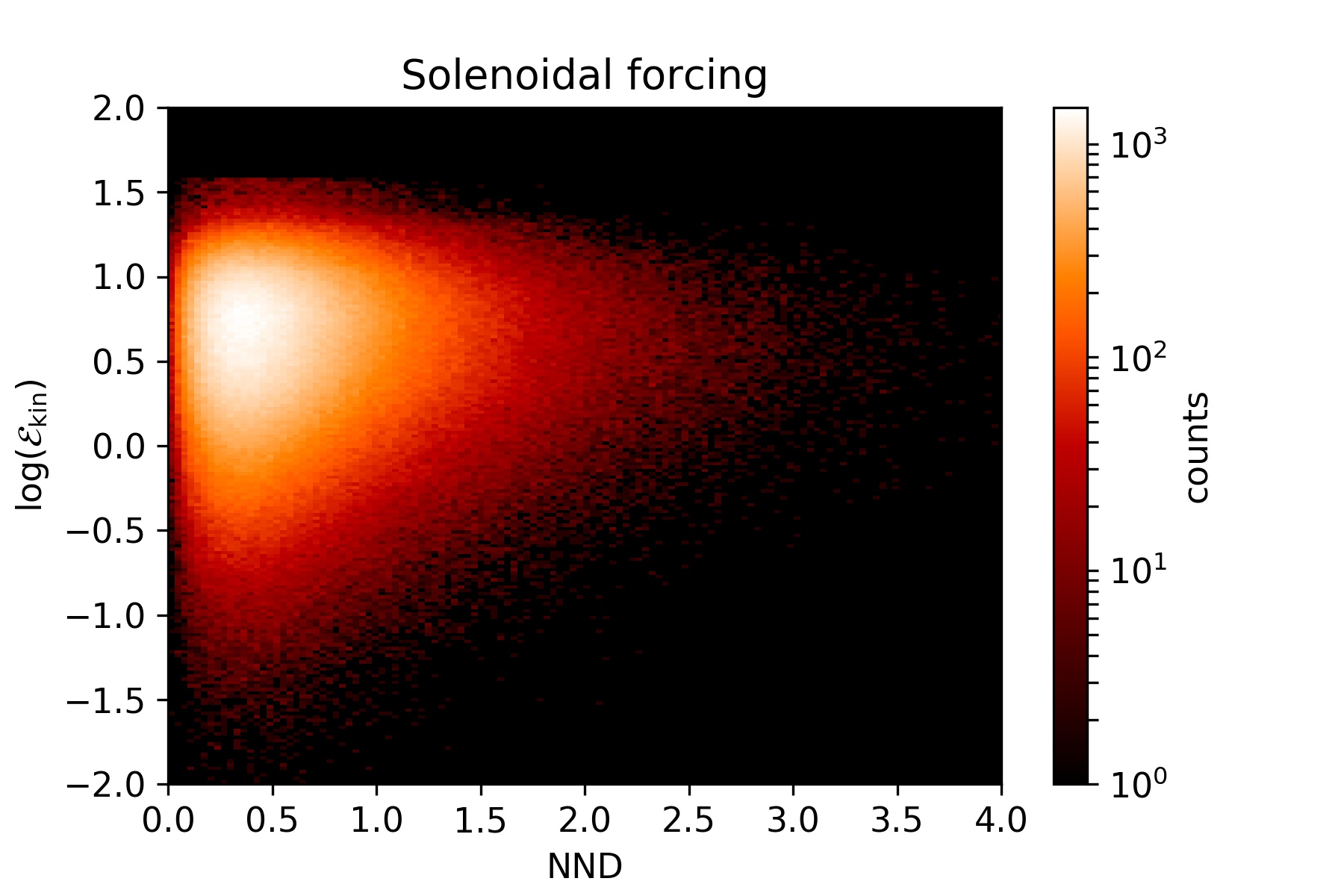}}
  \caption{\label{Ekin} Distribution of kinetic energyes (in terms of the $\mathcal{E}_{\rm kin}$ parameter) and first nearest neighbour distances (1-NND) for dust particles with maximal clustering ($\alpha = 0.033$, $a \approx 25$~nm) in the simulations. The left panel shows the simulation of compressive forcing, while the right panel shows the one with solenoidal forcing. Statistically, there are essentially no difference between the two cases.}
  \end{figure*}   
  
    \begin{figure*}
      \resizebox{\hsize}{!}{
      \includegraphics{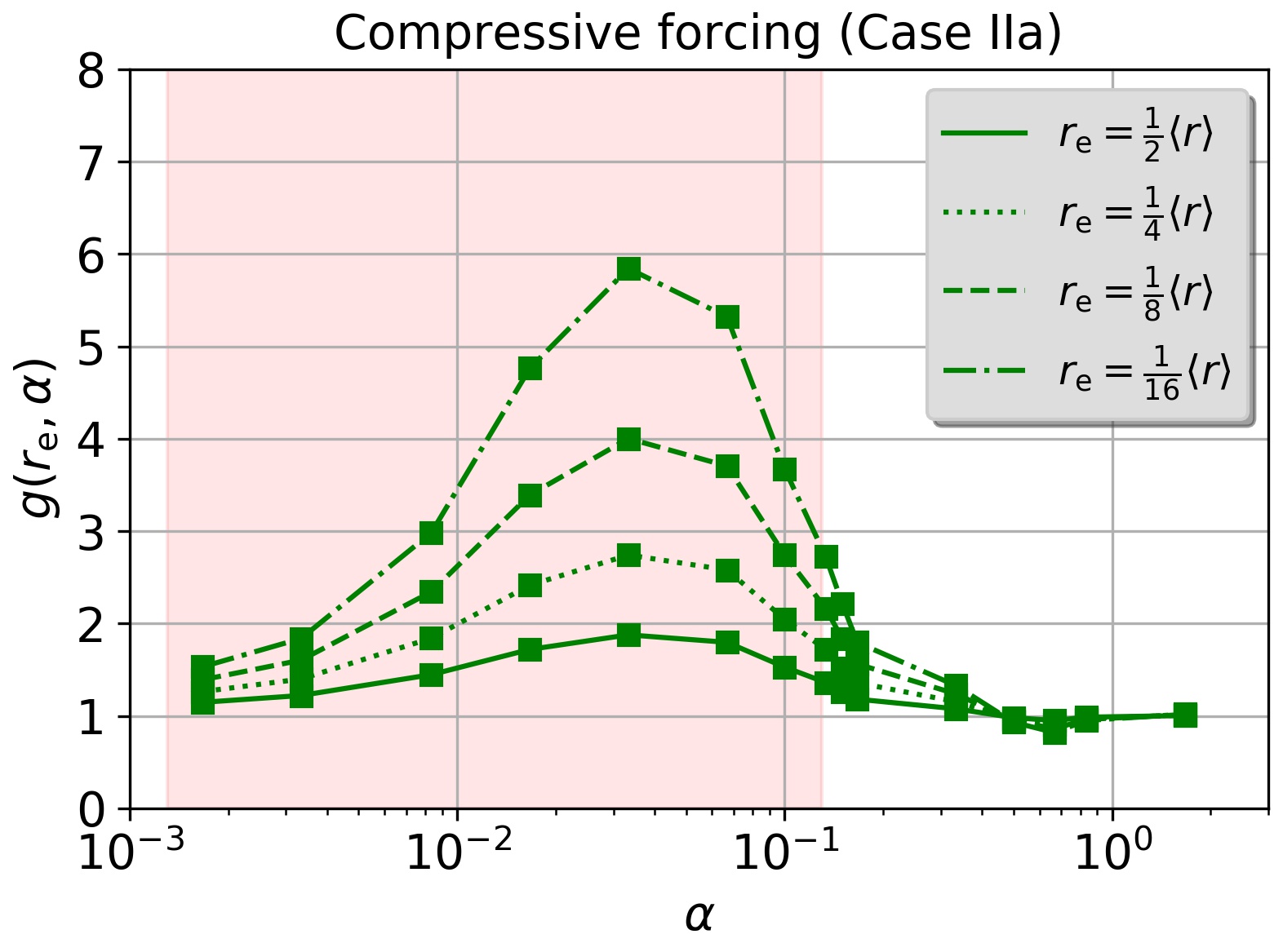}
      \includegraphics{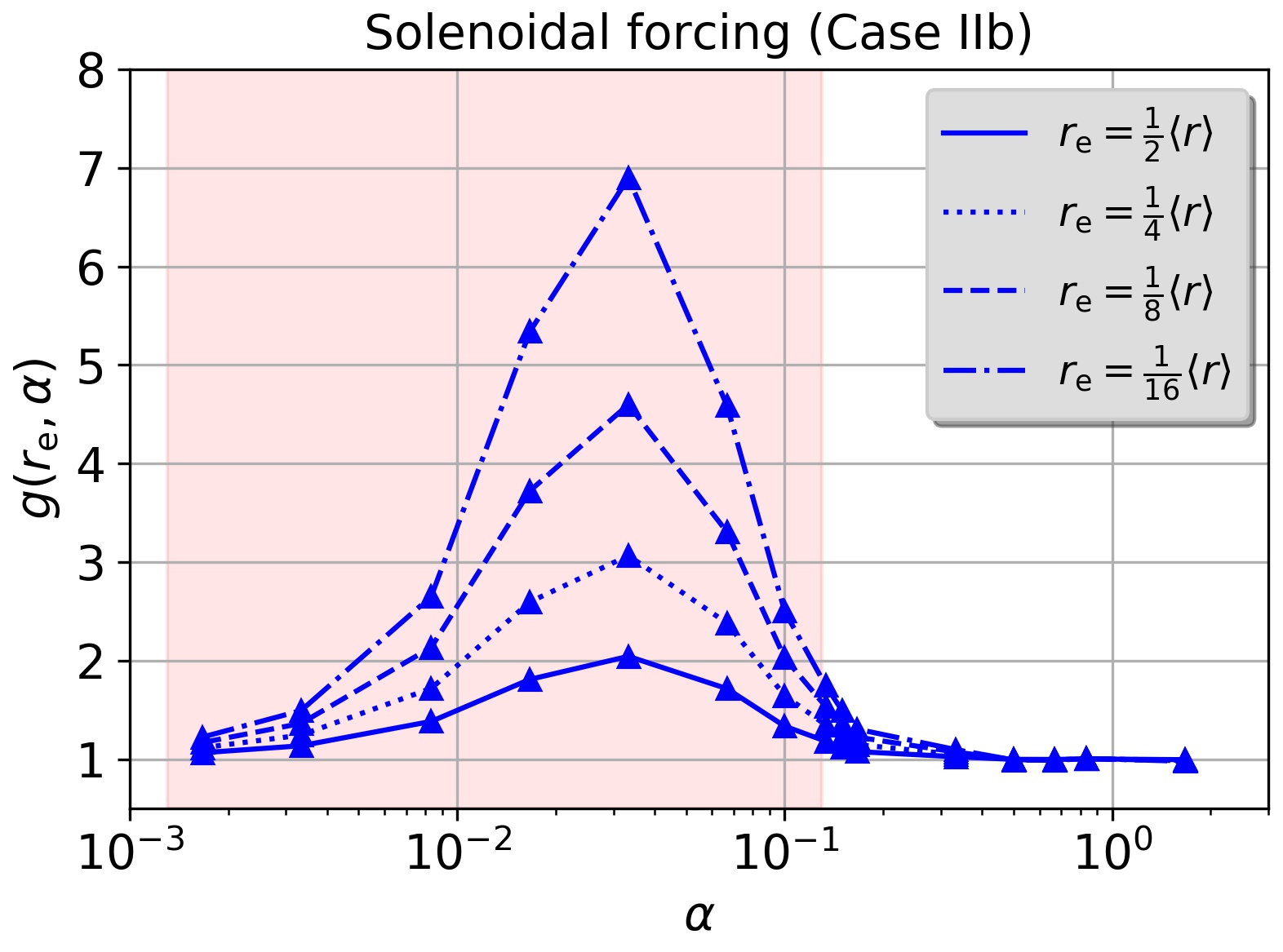}}
  \caption{\label{RDF} Radial distribution functions (RDFs) as functions of $\alpha$ for the cases of compressive (left) and solenoidal (right) stochastic forcing. The RDFs have been evaluated at four different separations $r_{\rm e}$ chosen to be ${1\over 2}$, ${1\over 4}$, ${1\over 8}$, and ${1\over 16}$ of the average 1-NND. 
   }
  \end{figure*}

  \begin{figure*}
      \resizebox{\hsize}{!}{
      \includegraphics{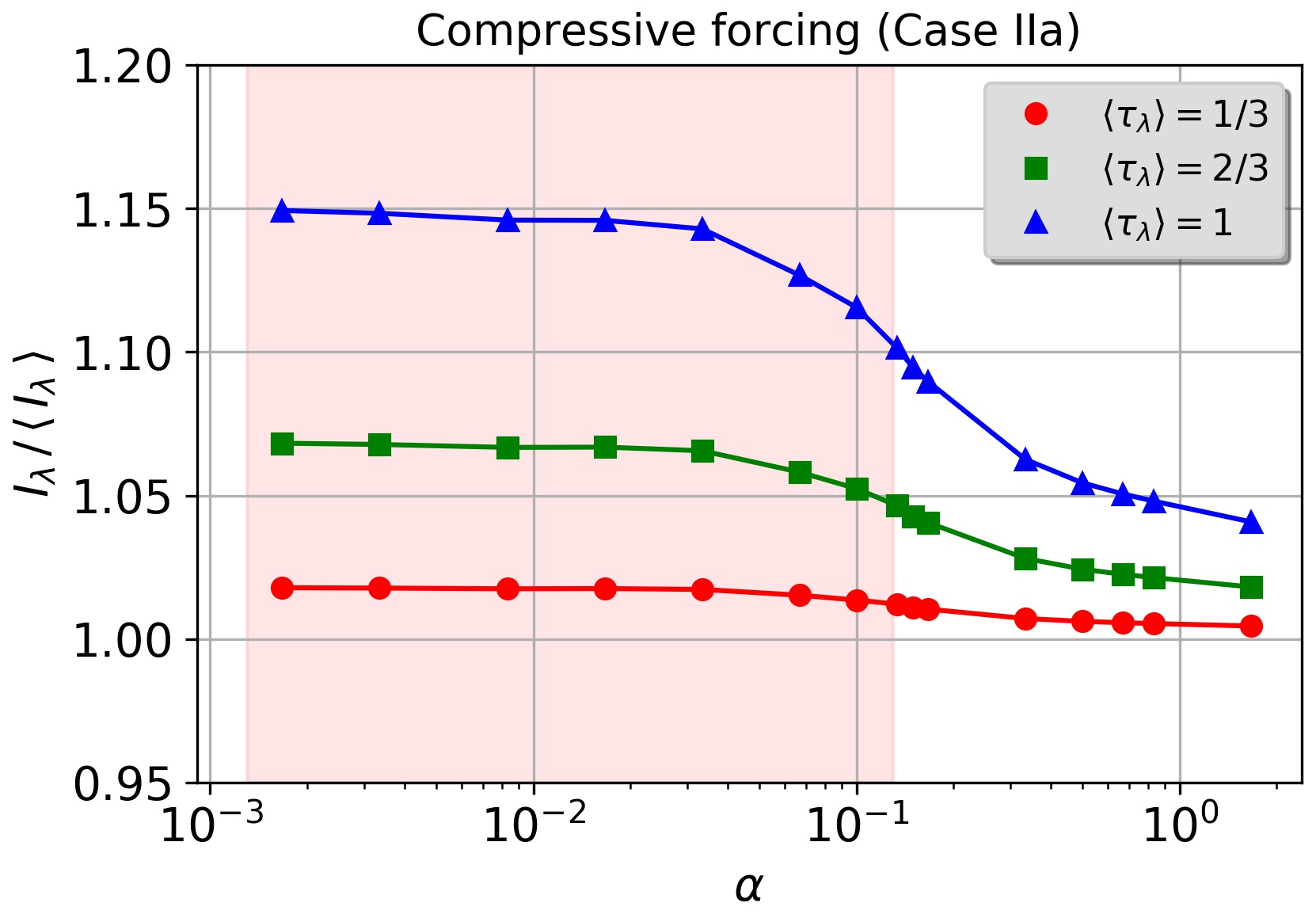}
      \includegraphics{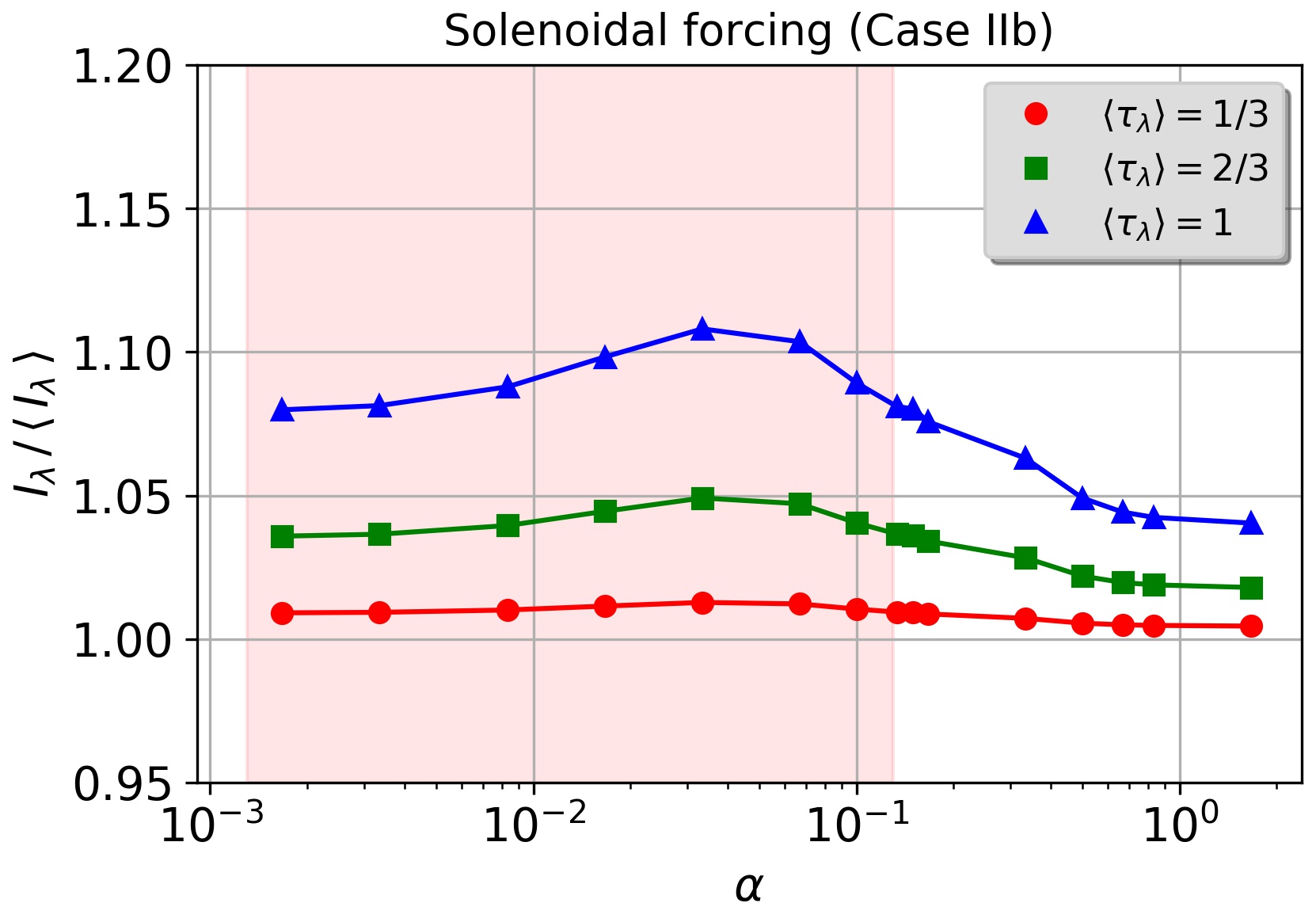}}
  \caption{\label{dustabs} Mean dust absorption as a function of grain size ($\alpha$) based on the distributions of maximally clustered grains ($\alpha = 0.033$, $a \approx 25$~nm) and assuming mean optical depth $\langle \tau_\lambda \rangle = 2/3$ for the cases of compressive (left) and solenoidal (right) stochastic forcing.}
  \end{figure*}    

  \begin{figure*}
            \resizebox{\hsize}{!}{
      \includegraphics{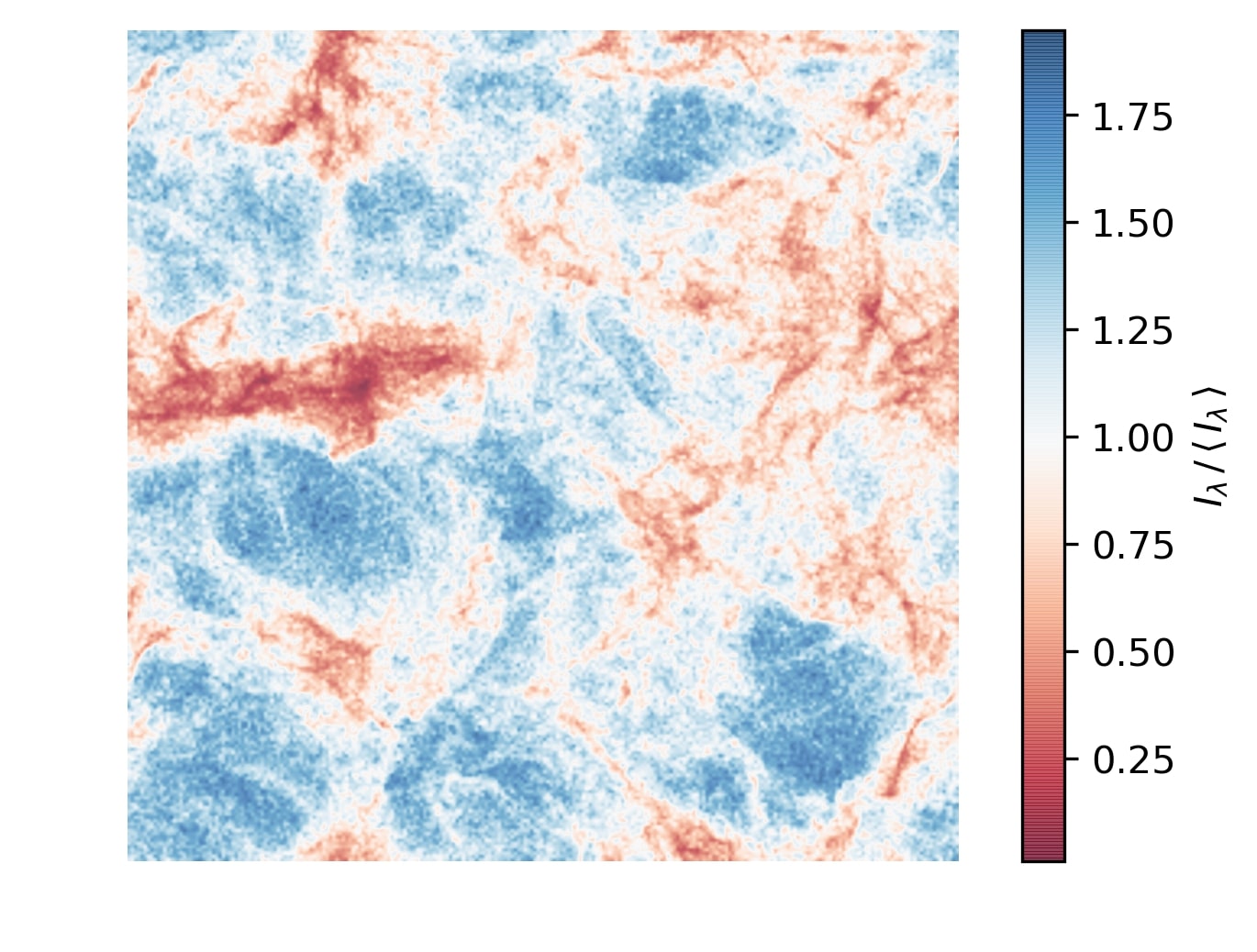}
      \includegraphics{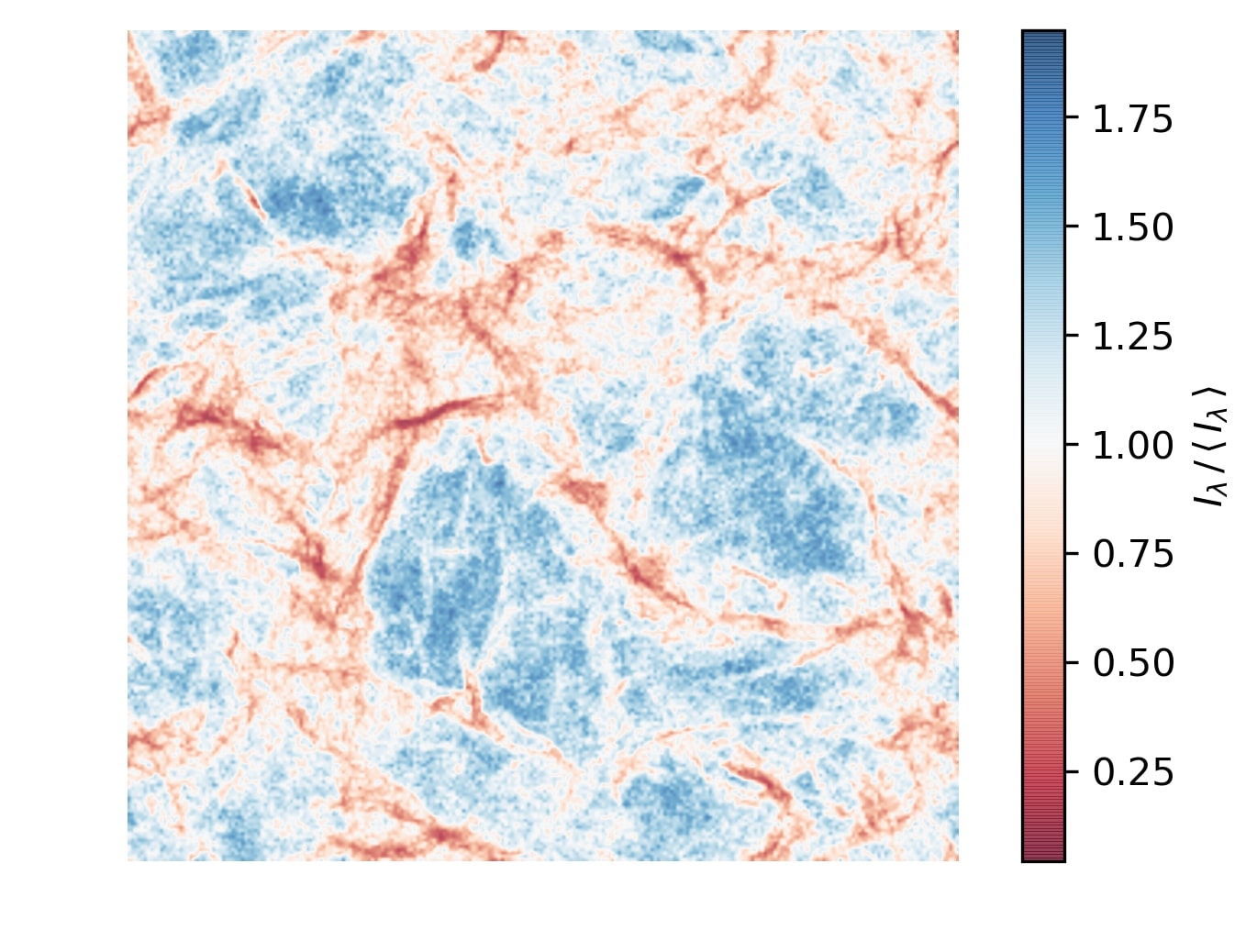}}
  \caption{\label{radtra} Dust absorption maps based on the distributions of maximally clustered grains ($\alpha = 0.033$, $a \approx 25$~nm) and assuming mean optical depth $\langle \tau_\lambda \rangle = 2/3$ for the cases of compressive (left) and solenoidal (right) stochastic forcing.}
  \end{figure*}

\section{Results and discussion}
\subsection{Clustering}
The resultant $d_2$ values are presented in Fig. \ref{D2}. One can clearly see that there does indeed exist a minimum in $d_2$ as a function of grain size also for supersonic compressible turbulence. However, there is one obvious qualitative difference: the ``low-$d_2$ valley'' is significantly wider and the $d_2$ minimum occurs at smaller $\alpha$ compared to a simulation of particles in nearly incompressible turbulence \citep{Bhatnagar18}, shown by grey diamonds connected by a dashed line in Fig. \ref{D2}\footnote{Note that in the paper by \citet{Bhatnagar18}, $d_2$ is presented as a function of ${\rm St}$. Their results were also obtained with the {\sc Pencil Code}, in which particles sizes are defined by $\alpha$ and the unit length and unit density of the simulation. \citet{Bhatnagar18} then converted their particle sizes into ${\rm St}$ for easier comparison with previous work on clustering of inertial particles in incompressible turbulence. In Figs. \ref{D2} and \ref{RANN} we have simply omitted this conversion.}. The clustering seems to begin at similar grains sizes ($\alpha \sim 1$), though. Very similar results have also been obtained by \citet[e.g.,][]{Hogan01,Bec03}. 

For the two different drag-force prescriptions we have tested, we note that the degrees of clustering are very similar. Case I yields a $d_2$ minimum at a somewhat lower $\alpha$, but since the values plotted in Fig. \ref{D2} are based on only a few simulation snapshots [except for the data taken from \citet{Bhatnagar18} which is based on a very long time series] the uncertainties in the $d_2$ values are comparable to the overall difference between the two cases. Furthermore, we note that the hypothesis described in Section \ref{dustflow}, i.e., that $\mathcal{W}_{\rm s} \ll 1$ because nano-dust grains stay coupled to the gas flow, on average, cannot be entirely correct. This is because the grains can briefly decouple from the flow so that the $\mathcal{W}_{\rm s} \gg 1$, even if just momentarily. The correction for $\mathcal{W}_{\rm s} \gg 1$ ends up playing a small, but not negligible, role in the grain dynamics and small-scale clustering.

The average nearest neighbour ratio, $R_{\rm ANN}$, is defined as the average of the ratio between the measured 1-NND and the corresponding quantity for a uniform, isotropic random distribution of particles (Poisson process). $R_{\rm ANN}$is a measure of both clustering and compression the same time and minima in $R_{\rm ANN}$ and $d_2$ occur at similar $\alpha$ (compare Figs. \ref{D2} and \ref{RANN}).  By considering $R_{\rm ANN}$ for non-inertial (``tracer'') particles, the effects of compressional increase of $n_{\rm d}$ can be isolated. In Fig. \ref{RANN} one can see that $R_{\rm ANN} < 1$ for small $\alpha$ and the tracer-particle runs showed that $R_{\rm ANN} \sim 0.8$ even though $d_2 = 3$. 

\subsection{Non-inertial particles and compression}
Although our main objective here is to study clustering of dust grains, we have also included two simulations (IIIa and IIIb) with non-inertial particles. Non-inertial particles have to be handled in separate runs due to how the {\sc Pencil Code} is structured. 

Mass-less particles is the limiting case of very small dust grains. Without drag forces, the particles will trace the gas flow and the distance between particles can only be reduced by compression of the gas in which they reside. An incompressible flow will leave the initial $R_{\rm ANN}$ unchanged, i.e., $R_{\rm ANN}=1$ if the initial condition is a homogeneous distribution. Our simulations of supersonic compressible turbulence yields a different result, however. 

Due to shock compression the gas forms local high-density regions, which is where the vast majority of the tracer particles end up. As a result, $R_{\rm ANN} \approx 0.8$, while the correlation dimension approaches $d_2 = 3$ as in incompressible turbulence. That is, the particles are inhomogeneously distributed, but they are {\it not fractally clustered}. The lower values of $R_{\rm ANN}$ is, in fact, a measure of the overall level of shock compression in the simulations and it is reasonable to think that for non-inertial particles, $R_{\rm ANN}$  will scale with $\mathcal{M}_{\rm rms}$, but since we have only considered a narrow range of $\mathcal{M}_{\rm rms}$ values (see Table \ref{simulations}), we cannot test this hypothesis here.

\subsection{Spatial distribution of kinetic energy}
To study how the kinetics and clustering of grains are connected, we consider the dimensionless specific kinetic energy for dust grains, i.e.,
\begin{equation}
\mathcal{E}_{\rm kin} = {1\over 2} \left({|\mathbfit{v}|\over c_{\rm s}}\right)^2.
\end{equation}
Comparing Figs. \ref{3Ddust_cmp} (compressive forcing) and \ref{3Ddust_sol} (solenoidal forcing) it seems that the most clustered grains have the lowest kinetic energies. There is also a qualitative difference between the types of forcing considered. Statistically the cases are actually very similar, which is apparent from Fig. \ref{Ekin}. We see that the highest $\mathcal{E}_{\rm kin} $ occur in particles with small separations (short 1-NND) and it is also evident that ``lonely grains'' typically do not have high $\mathcal{E}_{\rm kin}$.

The $\mathcal{E}_{\rm kin}$ distribution $P(\mathcal{E}_{\rm kin} )$ is created by a more or less stochastic process, where grains are ejected from vortices at arbitrary directions and positions and may hit other vortices, with different rotation, and thus be decelerated or accelerated, as well ejected again in another direction. Snapshots from the simulations show particles accumulating in the convergence zones between vortices having lost (or not gained very much) kinetic energy. The highest $\mathcal{E}_{\rm kin}$ occur in grains that were recently ejected from vortices. Isolated dust grains have generally rather average $\mathcal{E}_{\rm kin}$ values (see Fig. \ref{Ekin}), which is ar result of the stochasticity of the acceleration. Other grains with average $\mathcal{E}_{\rm kin}$ (which make up the vast majority of grains) are found in, or near, the convergence zones as a result of the same mechanisms; grains entering vortices will only stay there for a limited time \citep{Bhatnagar16} and this time can be regarded as a random variable.

\subsection{Remarks on grain-grain interaction and the increased probability of aggregation}
A majority of the dust grains populate regions which correspond to a total volume smaller than the whole simulation box, i.e., the average amount of space in between dust grains is reduced. This effect can be quantified by the RDF, $g(r,\alpha)$, which is peaking at the $\alpha$ where the $R_{\rm ANN}$ has its minimum (cf. Figs. \ref{RDF} and \ref{RANN}). The amplitude of $g(r,\alpha)$ depends on the radius $r$ and the implied scaling relation is $g \sim r^{d_2-3}$ as expected for small $r$. The amplitude at the peak is also somewhat ($\sim 25\%$) higher in Case IIb compared to Case IIa, but otherwise the simulation with compressive forcing gives a similar result compared to the solenoidal case. 

The frequency of collisions between two different particle species $i$ and $j$ is
\begin{equation}
\label{fcoll}
\varphi_{ij}\sim {\pi (a_i+a_j)^2 \,n_i \,n_j\,\Delta v_{ij}} = n_i \,n_j\,C_{ij},
\end{equation}
where $\Delta v_{ij}$ is the absolute velocity difference between the colliding particles. The average collision kernel $\langle C_{ij} \rangle$ is often approximated with $\langle C_{ij} \rangle \approx 2\pi\, (a_i+a_j)^2 c_{\rm s}\,\langle \mathcal{W} \rangle $, where $\langle \mathcal{W}\rangle$ require some kind of modelling of turbulence and drag \citep{Volk80}. Obviously,  $\langle \mathcal{W}\rangle$ does not include information about fractal clustering. By adding the RDF $g(r,a)$, we can account for fractal clustering. The collision kernel for identical particles $i$ then becomes $\langle C_{i}\rangle = 8\pi\,a_i^2 g(r_{\rm e}, a_i)\,c_{\rm s}\,\langle \mathcal{W}\rangle$ \citep[see, e.g.,][]{Wang00,Pan11,Pan14c}. In Fig. \ref{RDF} we show $g(r_{\rm e},\alpha)$ for the cases of compressive (left) and solenoidal (right) stochastic forcing, evaluated at four different radial separations $r_{\rm e}$ chosen to be ${1\over 2}$, ${1\over 4}$, ${1\over 8}$, and ${1\over 16}$ of the average NND $\langle r\rangle$ ($r_{\rm e} = {1\over 8}\langle r\rangle$ corresponds to the resolution limit set by the mesh used for computing the gas flow). The results shown in Fig. \ref{RDF} are qualitatively similar to those by \citet{Pan14c}. although the scaling is different. They also indicate that, for a given $\rho$, there is indeed a special particle size where the interaction frequency is higher than for any other size. For a gas density $\rho$ typical of an MC, the nano-dust scale is therefore of special interest.

In addition to clustering, $\varphi_{ij}$ also increases due to the turbulence component of $\Delta v_{ij}$ \citep[see, e.g.,][]{Ormel09}. Small grains, which couple well to the flow, will have small  $\Delta v_{ij}$, while decoupled large grains are not accelerated very much due to kinetic drag and therefore also have small  $\Delta v_{ij}$. That is, $g(r_{\rm e},\alpha)$ is large for small grains, but the collision frequency $\varphi_\mathrm{coll}$ is limited by  $\Delta v_{ij}$ and the largest increase of the probability of aggregation is obtained for accretion of grains with radii $a\sim 25-30\,$\AA~onto grains with $a\gtrsim 100\,${\AA}. This follows from the fact that at least one of the colliding grains must have a significant geometric cross-section, and both grains cannot be be tightly coupled to the flow, in order to achieve a high $\varphi_{ij}$ \citep{Bhatnagar18b}.

\subsection{Effects on dust absorption}  
To study the effects on nano-dust extinction, we consider a simplified RT problem of parallel rays of light incident on the simulation box, where we omit absorption and emission from the gas as described in Sect. \ref{sec:rte}. The intensity $I_{\lambda,\,i}$ of light surviving the passage through a column of the inhomogeneous distribution of a dust species $i$ relative to the mean intensity $\langle I_{\lambda,\,i}\rangle$ (taken of the whole projected area) is a function of the mean optical depth $\langle \tau_{\lambda,\,i} \rangle$ and the normalised number density of dust. 

An inhomogeneous distribution of dust leads to a higher throughput of photons. This effect depends on $\alpha$, but depends even more on  $\langle \tau_{\lambda,\,i} \rangle$ (see Fig. \ref{dustabs}). The fraction of surviving photons increases up to 15\% in Case IIa (compressive forcing) for $\langle \tau_\lambda\rangle = 1$ and peaks at about 10\% for Case IIb (solenoidal forcing). Apart from a somewhat lesser throughput, Case IIb is qualitatively different from Case IIa because of a maximum in ${I_{\lambda,\,i}/ \langle I_{\lambda,\,i}\rangle}$, which coincides with maximal clustering, i.e., the $\alpha$ range where $d_2$ and $R_{\rm ANN}$ have their minima. 

It is important to emphasise that the effect described above may depend on the geometrical depth of the modelled region. If we consider an elongated rectangular slab, instead of a box, where the integration of the RTE is done along the extended axis of the slab, the effective blocking of light will approach the uniform case (provided that the slab is sufficiently long).

\subsection{Nano grains and the extreme extinction in the sightline of GRB\,140506A}  
We will end this section by discussing an example of peculiar extinction, which may be due to clustered nano dust.
The afterglow of the gamma-ray burst GRB\,140506A display a very unusual extinction curve that was very steep in the ultraviolet \citep{Fynbo14}. This peculiar extinction was found to vary between two epochs separated by about 24 h. One possibility for this could be that the there is very strongly clustered dust in the foreground of the afterglow. The line-of-sight was also peculiar in other ways, indicating a very high density in the intervening material. \citet{Fynbo14} argued that GRB\,140506A  might be an example of an extreme 2175 \AA -feature, much wider and deeper than unusual. However, a follow-up analysis \citep{Heintz17} has revealed that the extreme 2175 \AA -feature scenario seems to be excluded by the data.

If extinction in the UV is dominated by nano-sized grains, which are very inhomogeneously distributed in the ISM, there ought to be extinction curves differing between objects separated by small angles on the sky. This variation should be stronger in the UV than at longer wavelengths as small grains dominate extinction at ultraviolet wavelengths. Concerning the spatial variations of the ultraviolet extinction curve the work of \citet{FP07} remains the most comprehensive. They report large line-to-sight variations and a lack of a correlation between variations in the ultraviolet and infrared parts of the extinction curve. 

\begin{figure}
      \resizebox{\hsize}{!}{
      \includegraphics{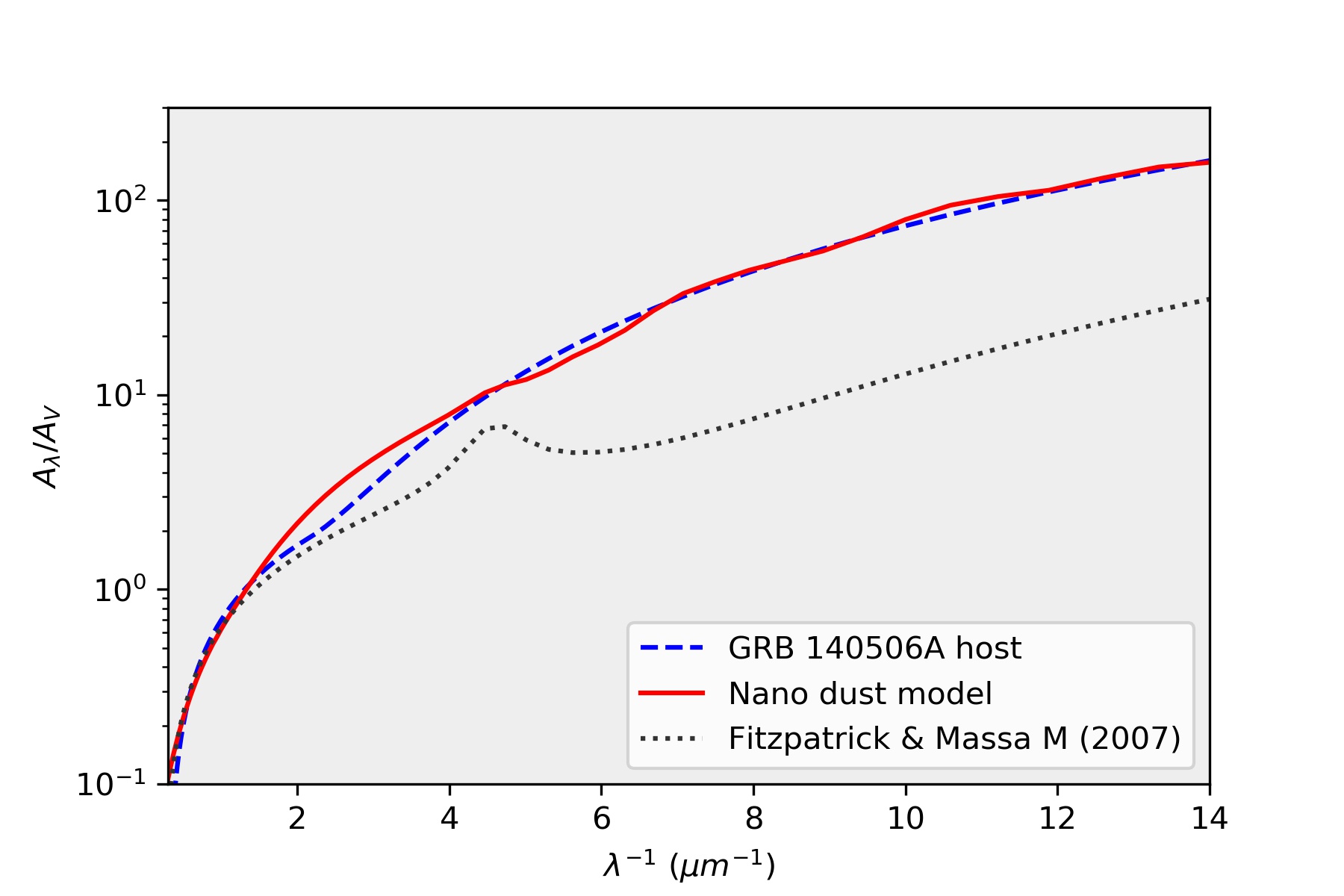}}
  \caption{\label{ecmodel} Nano dust model of the extinction towards GRB\,140506A. Note that the extinction curve for the GRB tend to coincide with that of the Galaxy at long wavelengths. The extreme UV extinction is in the nano dust model mainly due to tiny silicate grains.}
  \end{figure}

To test whether the extinction in the sightline of GRB\,140506A can be explained with clustered nano dust, we have made a simple extinction model based on silicates \citep{Draine01}, graphite \citep{Laor93} and metallic iron \citep{Palik91}, assuming all grains have a radius $a = 40$nm, which is in the middle of the ``clustering window'' ($a \sim 20-60$~nm) implied by the simulations presented above. We assume that the extinction is due to a high concentration of nano dust superimposed on an extinction law like the nominal \citet{FP07} curve, but scaled down to 25\% of the Galactic extinction. Thus, the nano dust is added like a local foreground screen on top of some generic host-galaxy extinction. A good fit to the exceptional extinction curve derived by \citet{Heintz17} is obtained with a nano-dust component consisting of 96\% silicate grains, 2\% graphite and 2\% metallic iron grains (see Fig. \ref{ecmodel}).

The extreme extinction in the sightline of GRB\,140506A is unusual and very rare. A plausible, although speculative, explanation might be that when an interstellar gas cloud with a high density of nano dust passes the sightline of the GRB, small-scale clustering may locally cause extreme extinction. The GRB\,140506A phenomenon should in such a scenario occur only very rarely and, depending on the distance between the GRB and the cloud, it may also explain fast variability of the extinction.

\section{Conclusions}
In the present paper we have investigated the small-scale clustering of nano-sized dust grains in an interstellar context using 3D high resolution ($1024^3$) periodic-boundary box simulations of stochastically forced, supersonic steady-state turbulence. The aim has primarily been to provide a bench mark for further study by establishing how clustering and concentration of particles in compressible hydrodynamic turbulence is different from to the incompressible case, which is much better understood. A fully realistic model of clustering of nano-grains in a turbulent MC must include charging of grains as well as magnetic fields, but this will have to wait until the purely kinetic picture is understood.

We conclude that kinetic drag on small particles in compressible turbulence show maximal clustering at a smaller particle size $\alpha$, and display a wider dip in the $d_2$--$\alpha$ relation around that size ($\alpha \approx 0.04$), compared to incompressible turbulence. Nano-dust grains must be regarded as inertial particles despite their tiny masses, which, among other things, is illustrated by the fact that $d_2 <3$ over the whole nano-dust range ($a = 1\dots100$~nm) assuming a scaling of the simulations which represents the conditions in an MC. 

The RDF $g(r,\alpha)$ has a maximum where $d_2$ has a minimum (at $\alpha \approx 0.04$), which corresponds to interstellar grains with radii $a \sim 20-60$~nm. The effective grain-number density increases by at least a factor 5--6 and the chance the collision kernel increases by a similar factor. Thus, the theoretical rate of nano-grain coagulation increases and the chance of building agglomerates starting from a population of mostly nano-sized grains without a preceding phase of grain growth by accretion of molecules goes from being almost impossible to at least plausible.

Simple radiative transfer through columns of the simulation box, show that the compression of the gas redistributes nano dust in such way that the total throughput of parallel rays of light incident on the box increases compared to the case of homogeneously distributed dust and gas. This effect has a clear dependence on the adopted mean optical depth for the dust; for $\langle \tau_\lambda\rangle = 2/3$ the increase of the throughput is 4--7\%, while it is 10--15\% for $\langle \tau_\lambda\rangle = 1$. There is also a dependence on $\alpha$ because this effect is only seen when grains couple well to the gas. In the case of solenoidally forced turbulence we also see a peak in the radiative throughput at $\alpha \approx 0.04$, where $d_2$ has its minimum (corresponding to maximal clustering). 

Finally, we note that extreme UV extinction, such in the sightline of GRB\,140506A \citep{Fynbo14, Heintz17},  could be the result of nano dust. An extinction model based on silicates, graphite and metallic iron, assuming grain sizes in the ``clustering window'' ($a \sim 20-60$~nm), appears to explain the broad and deep UV extinction feature of GRB\,140506A, and, possibly, its short-term variations too.

\section*{Acknowledgments}
This project is supported by the Swedish Research Council (Vetenskapsrådet), grant no. 2015-04505. The Cosmic Dawn center is funded by the DNRF. The anonymous reviewer is thanked for his/her insightful criticism, which helped to improve the manuscript. We also wish to thank our colleagues Akshay Bhatnagar and Dhrubaditya Mitra for sharing their simulation data and for all their help and support.

\bibliographystyle{mnras}
\bibliography{refs_dust}

\label{lastpage}
\end{document}